\PassOptionsToPackage{table}{xcolor}
\documentclass[sigconf,screen,authorversion,nonacm]{acmart}
\definecolor{BlockPurple}{HTML}{E4E1FA}
\usepackage{enumitem}
\usepackage{booktabs}
\usepackage{algorithm}
\usepackage{algpseudocode}
\usepackage{amsmath,amsthm}
\usepackage{graphicx}
\usepackage{tabularx}
\usepackage{array}
\usepackage{balance}
\usepackage{stfloats}
\usepackage{mdframed}
\usepackage{multirow}
\usepackage{needspace}
\AtBeginDocument{%
  }

\setcopyright{none}
\copyrightyear{2027}
\acmYear{2027}

\begin{document}

\title{Towards Efficient Reasoning in LLM-Based Recommender Systems via Model Merging}

\author{Linh Dieu Le}
\affiliation{%
  \institution{The University of Queensland}
  \city{Brisbane}
  \country{Australia}
}

\author{Tong Chen}
\affiliation{%
  \institution{The University of Queensland}
  \city{Brisbane}
  \country{Australia}
}

\author{Shazia Sadiq}
\affiliation{%
  \institution{The University of Queensland}
  \city{Brisbane}
  \country{Australia}
}

\author{Hongzhi Yin}
\affiliation{%
  \institution{The University of Queensland}
  \city{Brisbane}
  \country{Australia}
}

\author{Ming Jin}
\affiliation{%
  \institution{Griffith University}
  \city{Brisbane}
  \country{Australia}
}

\author{Junliang Yu}
\authornote{Corresponding author. Email: junliang.yu@griffith.edu.au}
\affiliation{%
  \institution{Griffith University}
  \city{Brisbane}
  \country{Australia}
}

\renewcommand{\shortauthors}{Le et al.}

\begin{abstract}
Large language model-based recommender systems are increasingly adopting slow-thinking models that generate step-by-step reasoning before making predictions, often achieving higher accuracy than fast-thinking models that predict directly. However, their reasoning traces are often unnecessarily verbose, increasing inference costs without commensurate accuracy gains. Existing training-based approaches to reasoning compression often incur substantial adaptation costs, while inference-time methods are brittle and difficult to scale. These limitations motivate model merging as a promising training-free direction for transferring specialised behaviours between models in a shared parameter space. In particular, merging a slow-thinking model with a fast-thinking counterpart provides a natural mechanism for balancing recommendation accuracy and reasoning conciseness. To this end, we propose, to our knowledge, the first model merging framework for reasoning compression in recommender systems. Unlike conventional merging methods that apply uniform merge coefficients across model components, our method performs fine-grained merging at the level of individual attention heads, capturing heterogeneous patterns in recommendation reasoning. Each attention head is assigned a distinct merge coefficient according to its contribution to critical reasoning evidence and its sensitivity to parameter change, enabling selective injection of the concise behaviour of the fast-thinking model into the slow-thinking model and reducing reasoning verbosity without compromising recommendation quality. Experiments on three benchmark datasets show that our method reduces reasoning length by up to 24.3\% while outperforming competitive model merging baselines in maintaining recommendation accuracy. The code is available at \url{https://github.com/linhledieu/REAM}.
\end{abstract}

\settopmatter{printacmref=false}
\renewcommand\footnotetextcopyrightpermission[1]{}
\maketitle

\begin{figure}[t]
    \centering
    \includegraphics[width=0.95\columnwidth]{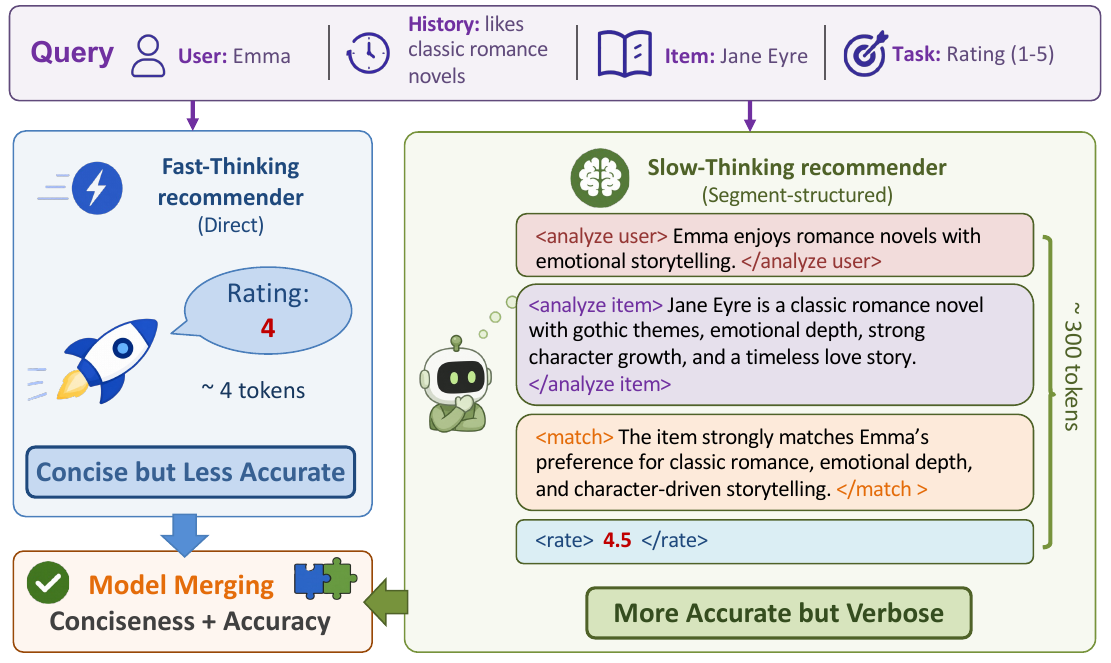}
    \caption{Motivation for merging fast- and slow-thinking recommenders toward concise and reliable prediction.}
    \label{fig:motivation}
    \vspace{-0.5em}
\end{figure}

\section{Introduction}
Large language models (LLMs) have become powerful foundations for recommender systems, enabling richer preference modelling from textual signals such as user histories, item descriptions, and reviews~\cite{lin2024recommendersystemsbenefitlarge, wu2024surveylargelanguagemodels}. Early LLM-based recommenders typically formulate recommendation as a direct prediction task, estimating user preferences from user--item context without explicit deliberate thinking, supporting fast inference~\cite{Bao_2023, liao2024llaralargelanguagerecommendationassistant, bao2023bistepgroundingparadigmlarge}. Building on these approaches, slow-thinking recommenders, echoing dual-process cognition~\cite{kahneman2011thinking}, incorporate Chain-of-Thought (CoT) reasoning to accumulate evidence before producing a rating~\cite{kong2025thinkrecommendationautonomousreasoningenhanced,fang2025reason4reclargelanguagemodels,Kim_2025,wei2023chainofthoughtpromptingelicitsreasoning}. Figure~\ref{fig:motivation} illustrates this distinction: fast-thinking recommendation maps the input directly to a rating, whereas slow-thinking recommendation first constructs a structured reasoning trace. This additional deliberation can improve accuracy, particularly when preference signals are sparse or implicit~\cite{li2026coldstart}, but often produces unnecessarily long traces regardless of input complexity, increasing inference latency and computational cost without commensurate benefits~\cite{sui2025stopoverthinkingsurveyefficient}.

Consequently, reducing reasoning verbosity in slow-thinking recommender systems remains a critical open challenge. Existing approaches to reasoning compression fall into two categories, but both face limitations in efficiency or robustness. Training-based methods, such as distillation and length-penalised optimisation, can shorten reasoning traces but require additional data construction and model adaptation, limiting their practical scalability~\cite{hou2025thinkprunepruninglongchainofthought,ma2025cotvalvelengthcompressiblechainofthoughttuning}. By contrast, inference-time methods avoid retraining by imposing token budgets or instructing the model to reason in shorter steps~\cite{han2025tokenbudgetawarellmreasoning,xu2025chaindraftthinkingfaster}, yet their effectiveness can vary across inputs. These limitations motivate model merging as a training-free alternative that transfers behaviours by combining models in parameter space, commonly through task vectors—the arithmetic differences between fine-tuned and pretrained weights~\cite{ilharco2023editingmodelstaskarithmetic}. Recommender systems provide a natural testbed for this approach: fast- and slow-thinking recommenders share the same preference-prediction objective but differ in the amount of reasoning generated before reaching a prediction~\cite{Bao_2023,kong2025thinkrecommendationautonomousreasoningenhanced}. Merging them may therefore transfer the concise generation behaviour of the fast-thinking model while preserving the reasoning required for accurate recommendation.

However, whether model merging can balance accuracy and efficiency in recommender systems remains unexplored, as prior work on reasoning compression has focused mainly on mathematical and coding tasks \cite{wu2025unlockingefficientlongtoshortllm, yao2025activationguidedconsensusmerginglarge}. Extending these methods to recommendation requires a more selective form of compression, since not all parts of a reasoning trace are equally safe to shorten: redundant elaboration may be reduced, whereas disrupting essential user--item evidence can directly alter the final prediction. Yet existing merging methods provide little control over how the reasoning process is altered. Wu et al.’s empirical analysis~\cite{wu2025unlockingefficientlongtoshortllm} shows that task-vector merging performance varies substantially with the choice of a single global coefficient. Seeking finer control, ACM~\cite{yao2025activationguidedconsensusmerginglarge} assigns a separate weight to each layer based on how similarly the two models behave, but its criterion does not explicitly account for how these updates reshape the reasoning trace. Taken together, existing work offers no reasoning-grounded basis for determining whether a shorter trace reflects useful compression or the loss of decision-critical evidence. Accordingly, to preserve recommendation accuracy, model merging must achieve selectivity at the component level, raising a central question: which components support the reasoning underlying the final prediction, and how should the fast-thinking update be allocated across them?

To determine where the fast-thinking update should be constrained, we first need a mechanistic signal of component-level reasoning importance. Prior work shows that reasoning-relevant behaviour is not distributed uniformly across the model but can be localised to a sparse subset of attention heads~\cite{voita2019analyzingmultiheadselfattentionspecialized,wu2024retrievalheadmechanisticallyexplains}. This pattern is especially pronounced in GRPO-trained slow-thinking models such as DeepSeek-R1~\cite{Guo_2025}, where reasoning-critical behaviour may emerge within only a small set of heads~\cite{park2026thinkingsparksemergentattention}. Together, these findings identify the attention head as the appropriate unit of analysis: heads serve distinct functional roles, and a small subset can contribute disproportionately to structured reasoning. At this granularity, a head’s importance can be inferred from how it governs information flow during generation, determining which parts of the input and preceding trace shape the model’s next output. Accordingly, this importance is especially consequential in slow-thinking recommender systems, where heads determine which user--item evidence enters the developing reasoning trace and which parts of that trace ultimately shape the final rating.

Building on this principle, we propose \textbf{REAM} (\textbf{R}easoning-H\textbf{E}ad-\textbf{A}ware \textbf{M}erging), a model merging method for reasoning compression in recommender systems. REAM formulates merging as selective transfer: it imports conciseness from the fast-thinking model while preserving the reasoning behaviour that supports the slow-thinking model's predictions. Rather than applying the fast-thinking update indiscriminately, REAM assigns each attention head a separate coefficient based jointly on its reasoning importance and sensitivity to parameter change. This constrains updates to heads critical for reasoning and prediction quality, while allowing less reasoning-critical and less sensitive heads to absorb more of the fast-thinking update. The resulting model requires no additional training or changes to decoding.

Our main contributions are as follows:
\begin{itemize}[leftmargin=1.05em, labelsep=0.35em, itemsep=0pt, topsep=1pt]
\item We formulate reasoning compression in slow-thinking recommender systems as a selective merging problem, where the fast-thinking update is allocated according to the contribution of each model component to recommendation performance.

\item We introduce REAM, to our knowledge, the first model merging framework for reasoning compression in recommender systems. By operating at the level of individual attention heads, REAM selectively transfers fast-thinking behaviour while limiting disruption to the heads most important for recommendation quality.

\item Experiments on Yelp, Amazon Book, and Amazon Music datasets show that REAM can substantially reduce reasoning length by up to 24.3\% while maintaining recommendation accuracy.
\end{itemize}

\section{Related Work}
\subsection{LLM-based Recommender Systems}
LLM-based recommender systems have progressed from direct prediction and knowledge augmentation toward explicit reasoning. Early work such as P5~\cite{geng2023recommendationlanguageprocessingrlp} unified recommendation tasks within a text-to-text paradigm, while InstructRec~\cite{zhang2023recommendationinstructionfollowinglarge} formulated recommendation as instruction following, enabling models to adapt their predictions to task-specific instructions. Subsequent systems such as TALLRec~\cite{Bao_2023}, LLaRA~\cite{liao2024llaralargelanguagerecommendationassistant}, and BIGRec~\cite{bao2023bistepgroundingparadigmlarge} further adapted general-purpose language models to recommendation-specific settings, improving preference prediction across diverse tasks. Alongside these LLM-centric approaches, a complementary line augmented conventional recommender systems with LLM-derived knowledge or collaborative signals~\cite{sun2024largelanguagemodelsenhanced}. Despite these advances, these methods struggle in cold-start settings, where limited interaction histories provide insufficient grounding for accurate recommendations~\cite{li2026coldstart}. To address this limitation, recent work introduces explicit reasoning before prediction, enabling models to infer latent relationships between user preferences and item characteristics. For instance, RDRec~\cite{wang2025rdrecrationaledistillationllmbased} distils CoT rationales from review text; EXP3RT~\cite{Kim_2025} and Reason4Rec~\cite{fang2025reason4reclargelanguagemodels} perform multi-step reasoning over user preferences and item attributes; and RecZero~\cite{kong2025thinkrecommendationautonomousreasoningenhanced} demonstrates that structured recommendation reasoning can emerge through reinforcement learning (RL) alone. Collectively, these methods suggest that explicit reasoning is particularly valuable for inferring preferences from sparse or implicit user--item signals, but often produces unnecessarily long traces and exacerbates overthinking~\cite{sui2025stopoverthinkingsurveyefficient, zhang2025reasoningaware}.

\subsection{Efficient Reasoning for LLMs}
Efficiently shortening reasoning traces remains an open challenge. Training-based methods shorten reasoning via rationale distillation~\cite{wang2025rdrecrationaledistillationllmbased}, length-penalised fine-tuning~\cite{ma2025cotvalvelengthcompressiblechainofthoughttuning}, RL with explicit length rewards~\cite{hou2025thinkprunepruninglongchainofthought,aggarwal2025l1controllinglongreasoning}, or token-level pruning~\cite{xia2025tokenskipcontrollablechainofthoughtcompression}. Their efficiency gains, however, incur adaptation costs: distillation depends on labelled or teacher-generated rationales, while fine-tuning, RL, and pruning require further target-model optimisation. By contrast, inference-time methods---including token-budget-aware generation~\cite{han2025tokenbudgetawarellmreasoning}, chain drafting~\cite{xu2025chaindraftthinkingfaster}, and cognitive-inspired sketching~\cite{aytes2025sketchofthoughtefficientllmreasoning}---avoid retraining by shortening or restructuring reasoning during generation. However, they often depend on predefined budgets, prompting strategies, or reasoning formats, making their effectiveness sensitive to input and task difficulty. Within recommender systems, efficiency methods largely follow the training-based route but tend to replace explicit reasoning rather than compress it. SCoTER transfers reasoning patterns offline into a structure-preserving architecture that removes LLM inference at deployment~\cite{jiang2026scoterstructuredchainofthoughttransfer}, while LatentR$^3$ replaces CoT with RL-learned latent tokens~\cite{zhang2025reinforcedlatentreasoningllmbased}. Although effective in reducing inference cost, these approaches may degrade prediction quality, interpretability, and adaptation to user context~\cite{gao2026reinforcedpreferenceoptimizationreasoningaugmented,Kim_2025}.

\subsection{Model Merging}
Model merging transfers capabilities by combining model weights, often without additional training~\cite{yang2025modelmergingllmsmllms}. Early work focused on checkpoint averaging: Model Soups~\cite{wortsman2022modelsoupsaveragingweights} averaged checkpoints fine-tuned from a shared base, while Task Arithmetic~\cite{ilharco2023editingmodelstaskarithmetic} represented fine-tuning updates as composable task vectors. To reduce interference between task vectors, TIES-Merging~\cite{yadav2023tiesmergingresolvinginterferencemerging} trims small updates and resolves sign conflicts, whereas DARE~\cite{yu2024languagemodelssupermario} drops and rescales parameter deltas. Another line derives non-uniform coefficients to control each component's contribution: AIM~\cite{nobari2025activationinformedmerginglargelanguage} uses activation magnitudes to preserve important base-model weights, while Sens-Merging~\cite{liu2025sensmergingsensitivityguidedparameterbalancing} adjusts coefficients according to parameter sensitivity within and across tasks. For reasoning compression, L2S-Merge~\cite{wu2025unlockingefficientlongtoshortllm} combines slow- and fast-thinking models to shorten responses while preserving accuracy, while ACM~\cite{yao2025activationguidedconsensusmerginglarge} pursues the same objective through layer-wise coefficients derived from activation statistics. Other work targets different objectives: RCP-Merging~\cite{yang2026rcpmergingmerginglongchainofthought} preserves long-CoT capability when integrating domain-specific knowledge, whereas RAIN-Merging~\cite{huang2026rainmerginggradientfreemethodenhance} uses variation in attention to instruction-relevant tokens to guide coefficient selection for instruction following. Model merging has also been explored in recommender systems, but for different objectives. An early work used Fisher-weighted merging to combine contrastive sequential recommenders~\cite{ryu2023fisherweightedmergecontrastivelearning}, while more recent work merges models for multi-domain, cross-domain, federated, and generative recommendation~\cite{yi2025multimodalrecipe,hou2025weaverecllmbasedcrossdomainsequential,chen2025breakingaggregationbottleneckfederated,kim2026mergerec,wei2026mmgridnavigatingtemporalawarecrossdomain}. These methods consolidate recommendation knowledge rather than compressing explicit reasoning.

\section{Preliminaries}
\subsection{Task Setting}
In this work, we study model merging for rating prediction in recommender systems. Let $\mathcal{U}$ and $\mathcal{I}$ denote the sets of users and items. Given a user $u \in \mathcal{U}$, their interaction history, and a target item $i \in \mathcal{I}$, the goal is to predict the rating $\hat{r} \in \mathbb{R}$ that $u$ would assign to $i$. Unlike direct-prediction LLM recommenders~\cite{Bao_2023,liao2024llaralargelanguagerecommendationassistant,bao2023bistepgroundingparadigmlarge}, slow-thinking recommenders first generate an explicit reasoning trace before producing $\hat{r}$~\cite{fang2025reason4reclargelanguagemodels,kong2025thinkrecommendationautonomousreasoningenhanced,Kim_2025}. This intermediate trace introduces overhead, requiring a long sequence before the final rating. 

Following RecZero~\cite{kong2025thinkrecommendationautonomousreasoningenhanced}, we organise the generated output into four sequentially conditioned segments: \texttt{<analyze\_user>} for user-preference analysis, \texttt{<analyze\_item>} for item-characteristic analysis, \texttt{<match>} for user--item compatibility reasoning, and \texttt{<rate>} for the final numerical prediction. These form the \textsc{user}, \textsc{item}, \textsc{match}, and \textsc{rate} segments; the first three constitute the reasoning trace, while $\hat{r}$ is generated in \textsc{rate} conditioned on it. This structure is adopted for its discrete regions, reflecting staged reasoning in recent slow-thinking recommenders~\cite{fang2025reason4reclargelanguagemodels,Kim_2025}; REAM adapts to other formats by redefining these regions accordingly (Appendix~\ref{app:template_generality}).

\subsection{Task Vectors and Model Merging}
A \emph{task vector}~\cite{ilharco2023editingmodelstaskarithmetic} is the weight difference $\Delta=\theta-\theta_B$ between a fine-tuned model $\theta$ and a base model $\theta_B$ of the same architecture, so parameters occupy identical positions. Task vectors relative to the same reference can therefore be combined as
\[
\theta_{\mathrm{merged}}
=
\theta_B+\sum_{m=1}^{M}\lambda_m\Delta_m,
\]
where $\Delta_m$ is the update for the $m$-th model and $\lambda_m\in[0,1]$ controls its contribution. We instantiate this framework with two recommendation models represented relative to the same reference $\theta_B$:
\begin{itemize}[leftmargin=*, itemsep=1pt, topsep=3pt]
    \item \textbf{Slow-thinking model} $\theta_S$ (RecZero~\cite{kong2025thinkrecommendationautonomousreasoningenhanced}): generates three structured reasoning segments followed by a \textsc{rate} segment. Its task vector is $\Delta_S=\theta_S-\theta_B$.
    \item \textbf{Fast-thinking model} $\theta_F$ (TALLRec~\cite{Bao_2023}): predicts ratings without intermediate reasoning. We adapt TALLRec from binary like/dislike prediction to direct rating prediction and fully fine-tune it so that its parameter differences are expressed over full weight matrices, matching our merging granularity (Appendix~\ref{app:tallrec_adaptation}). Its task vector is $\Delta_F=\theta_F-\theta_B$.
\end{itemize}
The two models share the same rating-prediction objective but differ in how they reach a prediction (see Appendix~\ref{app:format} for examples). Section~\ref{sec:method} specialises this formulation by using $\theta_S$ as the anchor and partitioning the fast-thinking update into components indexed by $b$, with $\alpha_b$ controlling how much of each component is merged.

\section{Proposed Method}
\label{sec:method}
\begin{figure*}[t]
    \centering
    \includegraphics[width=\textwidth]{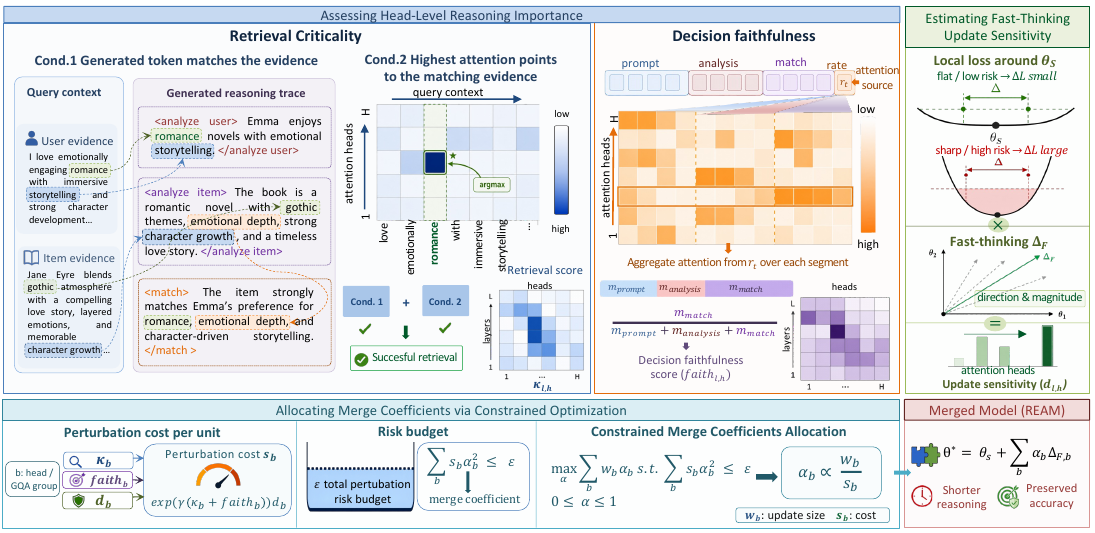}
    \caption{Overview of the proposed framework.}
    \vspace{-1em}
    \label{fig:framework}
\end{figure*}

To transfer concise generation behaviour from a fast-thinking recommender to a slow-thinking one while preserving the reasoning that supports its predictions, we develop our method by investigating three questions (Figure~\ref{fig:framework} summarises the resulting pipeline): (1) which attention heads are most important to the reasoning underlying the final prediction (\S\ref{sec:head_risk})? (2) How sensitive is each head to the fast-thinking update (\S\ref{sec:head_update_sensitivity})? (3) How should the update be allocated to improve conciseness while limiting degradation in reasoning and recommendation quality (\S\ref{sec:allocation})? 

\subsection{Assessing Head-Level Reasoning Importance}
\label{sec:head_risk}

Before allocating the fast-thinking update, we first assess each head’s importance to the reasoning underlying the final prediction, since reasoning-critical heads may require more conservative updates to preserve recommendation quality. Since attention heads govern information flow during generation, their importance may differ across reasoning stages. Inspired by Wu et al. \cite{wu2024retrievalheadmechanisticallyexplains}, we posit that a head’s importance may be reflected in its ability to retrieve relevant user--item evidence from the input context and previously generated trace to support the developing reasoning; we define this importance as \emph{retrieval criticality}. Later, at the \textsc{rate} step, we posit that its importance may instead arise from connecting the final prediction to the preceding reasoning, attending to key conclusions established in the trace; we refer to the strength of this connection as \emph{decision faithfulness}. These two complementary signals can serve as proxies for each head's reasoning importance.

\subsubsection{Retrieval Criticality.}
We operationalise retrieval criticality by measuring how frequently each head retrieves segment-relevant evidence throughout the reasoning trace. Frequent retrieval indicates that the head repeatedly restores information needed by subsequent reasoning stages, whereas disrupted retrieval can cause later generation to drift from the user--item context and propagate errors to the final prediction.

Concretely, we follow the retrieval-head criterion of Wu et al.~\cite{wu2024retrievalheadmechanisticallyexplains}, which identifies retrieval by aligning the generated token with the source position receiving a head's strongest attention. Since slow-thinking recommendation traces comprise distinct segments, we evaluate this criterion separately within each one. For an example $x$, consider head $h$ in layer $\ell$, denoted $(\ell,h)$, at decoding step $t$ of reasoning segment $s$. We count retrieval as successful only if both conditions hold: (Cond.1) the generated token $w_t$ matches a token in the segment-relevant evidence region $K^{(s)}(x)$; and (Cond.2) head $(\ell,h)$ assigns its maximum attention to a position containing a matching token within that region. Here, $K^{(s)}(x)$ is defined by the role of segment $s$. For \textsc{user} and \textsc{item}, it contains the corresponding user or item context. For \textsc{match}, it also includes the preceding \textsc{user} and \textsc{item} analyses, since compatibility reasoning should integrate the original user--item evidence with conclusions formed earlier.

We aggregate successful retrieval events over the slow-thinking calibration set $\mathcal{D}_{\mathrm{slow}}$, comprising 500 correctly predicted, well-formed prompt--trace pairs from the validation split (stability analysis in Appendix~\ref{app:calibration_stability}). To reduce spurious matches, stopword and punctuation steps are excluded before evaluating Cond.1--2, and successful retrievals are TF--IDF-weighted to emphasise rare tokens.
\begin{equation}
R_{\ell,h}
=
\frac{
\sum_{x \in \mathcal{D}_{\mathrm{slow}}}
\sum_{s \in \mathcal{S}}
\sum_{t}
\tau_x(w_t)\cdot g_{\ell,h,t}^{(s)}(x)
}{
\sum_{x \in \mathcal{D}_{\mathrm{slow}}}
\sum_{s \in \mathcal{S}}
|K^{(s)}(x)|
+\delta
},
\label{eq:retrieval_criticality}
\end{equation}
where $\mathcal{S}=\{\textsc{user},\textsc{item},\textsc{match}\}$, and $t$ ranges over content-token steps in segment $s$; $g_{\ell,h,t}^{(s)}(x)\in\{0,1\}$ indicates whether head $(\ell,h)$ satisfies Cond.1--2 at step $t$; $\tau_x(w_t)$ is the token's TF--IDF weight (Appendix~\ref{app:tfidf_weighting}); $|K^{(s)}(x)|$ is the evidence region's content-token count for segment $s$; and $\delta$ prevents division by zero. Thus, $R_{\ell,h}$ measures informativeness-weighted retrieval success per evidence-region content token, aggregated over $\mathcal{D}_{\mathrm{slow}}$.

However, $R_{\ell,h}$ is unbounded and strongly right-skewed, with a small subset of heads accounting for most retrievals. We map it to a bounded, comparable weight via a log--z-score--sigmoid transform, $\kappa_{\ell,h}=\sigma\!\left((\log(R_{\ell,h}+\delta)-\mu_R)/s_R\right)$, where $\sigma(\cdot)$ is the logistic sigmoid and $\mu_R,s_R$ are the mean and standard deviation of $\log(R_{\ell,h}+\delta)$ across heads. Intuitively, larger $\kappa_{\ell,h}$ indicates greater retrieval criticality.

\subsubsection{Decision Faithfulness.}
The performance of slow-thinking recommenders depends not only on preserving relevant information throughout the trace, but also on whether the final prediction faithfully draws on the preceding reasoning. This raises a complementary question: which heads make the key conclusions visible to the \textsc{rate} step, guiding the model toward the correct rating rather than attending broadly across the context? We quantify this role by measuring how strongly each head connects the final rating to the part of the preceding reasoning that most directly supports it.

Since the reasoning trace comprises segments, this measure requires identifying which segment the rating relies on most. We evaluate this on $\mathcal{D}_{\mathrm{slow}}$ by replacing \textsc{user}, \textsc{item}, and \textsc{match} with length-matched filler text and measuring the Jensen--Shannon divergence between original and perturbed rate-token distributions. Across three datasets, replacing \textsc{match} yields a divergence 3.3--5.1$\times$ larger than replacing \textsc{user} or \textsc{item}, indicating that the rating is most sensitive to perturbations of the \textsc{match} segment (Appendix~\ref{app:segment_jsd}). Accordingly, we measure decision faithfulness as the share of \textsc{rate}'s attention directed to \textsc{match}, relative to the other regions.

To compute this signal, we partition the sequence preceding \textsc{rate} into three regions: \textsc{prompt}, containing the input context; \textsc{analyze}, combining \textsc{user} and \textsc{item}, both of which affect the rating distribution substantially less than \textsc{match} (Appendix~\ref{app:segment_jsd}); and \textsc{match}. For each example $x \in \mathcal{D}_{\mathrm{slow}}$, let $T_{\textsc{rate}}(x)$ denote positions of rating-content tokens in the \textsc{rate} segment.
\begin{equation}
m_{\ell,h}^{r}
=
\frac{1}{|\mathcal{D}_{\mathrm{slow}}|}
\sum_{x \in \mathcal{D}_{\mathrm{slow}}}
\frac{1}{|T_{\textsc{rate}}(x)|}
\sum_{t \in T_{\textsc{rate}}(x)}
\sum_{\tau \in \Omega_r(x,t)}
\mathrm{Att}_{\ell,h}^{(x)}[t,\tau],
\end{equation}
where $\Omega_r(x,t)$ denotes positions in region $r$ before $t$, and $\mathrm{Att}_{\ell,h}^{(x)}[t,\tau]$ is the attention weight from $t$ to $\tau$. We use total attention mass, not per-token, to capture each region's contribution to the decision-time representation; Appendix~\ref{par:length-normalised-faith} tests length robustness.

We define decision faithfulness as the share of attention directed to \textsc{match} among the three tracked regions:
\begin{equation}
\mathrm{faith}_{\ell,h}
=
\frac{
m_{\ell,h}^{\textsc{match}}
}{
m_{\ell,h}^{\textsc{match}}
+
m_{\ell,h}^{\textsc{analyze}}
+
m_{\ell,h}^{\textsc{prompt}}
+
\delta
}.
\end{equation}
The score satisfies $\mathrm{faith}_{\ell,h}\in[0,1)$, where larger values indicate head $(\ell,h)$ directs a greater share of decision-time attention to the compatibility judgement in \textsc{match} than to the prompt or analyses.

It should be noted that both $\kappa_{\ell,h}$ and $\mathrm{faith}_{\ell,h}$ are defined at the query-head level. Under grouped-query attention (GQA), several query heads share key and value projections. Let $\mathcal{H}_{\ell,g}$ denote the set of query heads belonging to GQA group $g$ in layer $\ell$. We assign the shared projections the maximum score among their member heads:
\[
\kappa_{\ell,g}^{K/V}
=
\max_{h\in\mathcal{H}_{\ell,g}}\kappa_{\ell,h},
\qquad
\mathrm{faith}_{\ell,g}^{K/V}
=
\max_{h\in\mathcal{H}_{\ell,g}}\mathrm{faith}_{\ell,h}.
\]
This conservative aggregation treats the shared $K/V$ parameters as reasoning-critical whenever any associated query head is critical.

\subsection{Estimating Fast-Thinking Update Sensitivity}
\label{sec:head_update_sensitivity}

The preceding analysis characterises each head's importance to the reasoning underlying the final prediction, but not how safely its parameters can be modified. Around the trained slow-thinking solution, the loss may be flat along some parameter directions, tolerating larger changes with little effect, but sharp along others, where even small perturbations can substantially alter the model's behaviour. Since the fast-thinking update modifies each head along a specific direction and magnitude, its risk depends on the loss sensitivity along that update. Following Fisher-weighted merging~\cite{matena2022mergingmodelsfisherweightedaveraging}, we estimate the sensitivity of the slow-thinking model using the diagonal empirical Fisher of $\theta_S$. We then weight this sensitivity by the squared fast-thinking update to obtain a component-level risk measure, which we term \emph{head-level update sensitivity} (see Appendix~\ref{app:fisher_derivation} for the derivation).

For each calibration example $x \in \mathcal{D}_{\mathrm{slow}}$, let
$Y(x)=(y_1,\ldots,y_{|Y(x)|})$ denote its complete generated response. Because a head may influence token predictions throughout the response, we compute its sensitivity using the average teacher-forced loss over $Y(x)$:
\begin{equation}
\ell_{\theta_S}(x)
=
-\frac{1}{|Y(x)|}
\sum_{t=1}^{|Y(x)|}
\log p_{\theta_S}
\!\left(
y_t \mid y_{<t},\mathrm{prompt}(x)
\right),
\end{equation}
where $\mathrm{prompt}(x)$ denotes the prompt portion of $x$. Averaging over $|Y(x)|$ prevents longer responses from dominating the estimate. For each parameter $\theta^{(j)}_{S}$, we compute the diagonal empirical Fisher as
\begin{equation}
F_j
=
\frac{1}{|\mathcal{D}_{\mathrm{slow}}|}
\sum_{x \in \mathcal{D}_{\mathrm{slow}}}
\left(
\frac{\partial \ell_{\theta_S}(x)}{\partial \theta^{(j)}_{S}}
\right)^2 .
\end{equation}
To obtain head- and group-level sensitivities, we partition Fisher entries by the structure of the attention projection matrices $W_Q$, $W_K$, $W_V$, and $W_O$ (query, key, value, output). $W_Q$ and $W_O$ are divided by query head, while $W_K$ and $W_V$ are divided by shared GQA group. Under our convention, each head in $W_O$ corresponds to the columns of its output in the concatenated multi-head representation.

Let $c$ index a single-projection slice: a $Q$ or $O$ slice for query head $(\ell,h)$, or a $K$ or $V$ slice for GQA group $(\ell,g)$. We define its update sensitivity as the Fisher-weighted squared fast-thinking update:
\begin{equation}
d_c
=
\sum_{j \in c}
F_j
\left(\Delta_{F,j}\right)^2 .
\label{eq:fisher_sensitivity}
\end{equation}
A larger $d_c$ indicates loss sensitivity to that portion of the fast-thinking update, calling for more conservative merging (derivation in Appendix~\ref{app:head_update_sensitivity_derivation}). To match merge-coefficient granularity, we average slice sensitivities: $d_{\ell,h}^{QO}=(d_{\ell,h}^{Q}+d_{\ell,h}^{O})/2$ for query head $(\ell,h)$, and $d_{\ell,g}^{KV}=(d_{\ell,g}^{K}+d_{\ell,g}^{V})/2$ for GQA group $(\ell,g)$. Hereafter, $b$ indexes either a Q/O head or K/V group, and $d_b$ denotes its sensitivity.

\subsection{Allocating Merge Coefficients via Constrained Optimization}
\label{sec:allocation}

Together, reasoning importance and update sensitivity provide the reasoning-grounded basis for selective merging by indicating which heads should be protected and how cautiously each can be modified. We translate these signals into head-aware merge coefficients through a constrained optimisation that transfers as much of the fast-thinking update as possible while limiting the overall risk to recommendation-relevant reasoning.

Concretely, we combine retrieval criticality, decision faithfulness, and update sensitivity into a perturbation weight that represents the risk of perturbing unit $b$ along the fast-thinking update:
\[
s_b
=
\exp\!\left(\gamma(\kappa_b+\mathrm{faith}_b)\right)d_b,
\]
where $b$ indexes either the paired $Q/O$ parameters of a query head or the shared $K/V$ parameters of a GQA group. The exponential factor amplifies update sensitivity of reasoning-critical units, with $\gamma$ controlling strength (ablations: Appendix~\ref{app:s_b_form}; normalisation:~\ref{app:normalization_choices}). The allocation is robust to uniform-attention bias in $\mathrm{faith}_b$, since any constant offset rescales every $s_b$ equally (Appendix~\ref{par:baseline-invariance}).

Given these perturbation weights, we choose merge coefficients that maximise the total applied fast-thinking update while limiting aggregate reasoning-grounded risk. We solve this separately for the $Q/O$ and $K/V$ unit sets, $\mathcal{B}^{QO}$ and $\mathcal{B}^{KV}$, because they operate at different granularities under GQA and exhibit different update and risk scales. For either set $\mathcal{B}$,
\begin{equation}
\max_{\{\alpha_b\}_{b\in\mathcal{B}}}
\sum_{b\in\mathcal{B}} w_b\alpha_b
\quad
\mathrm{s.t.}
\sum_{b\in\mathcal{B}} s_b\alpha_b^2 \leq \varepsilon,
\qquad
0 \leq \alpha_b \leq \bar{\alpha},
\label{eq:budgeted_allocation}
\end{equation}
where $w_b=\|\Delta_{F,b}\|_F$ denotes the magnitude of the fast-thinking update for unit $b$, and $\bar{\alpha}$ caps each merge coefficient, mainly to prevent unstable extrapolation rather than as a sensitive hyperparameter (value in Appendix~\ref{app:experimental_configuration}); $\varepsilon$ controls the aggregate perturbation. We set $\varepsilon=\rho\bar{\alpha}^2\sum_{b\in\mathcal{B}} s_b$, where $\rho\in(0,1]$ specifies the permitted fraction of the perturbation incurred when $\alpha_b=\bar{\alpha}$ for all $b$.

Since the objective is linear and the constraint is convex quadratic for $s_b\geq0$, KKT conditions characterise the global optimum (Appendix~\ref{app:kkt_derivation}). When binding, they yield the water-filling solution
\[
\alpha_b^\star
=
\operatorname{clip}\!\left(
\frac{w_b}{2\mu s_b},
0,
\bar{\alpha}
\right),
\]
where $\mu>0$ is determined separately for the $Q/O$ and $K/V$ allocations by bisection. For units with $s_b=0$, we set $\alpha_b^\star=\bar{\alpha}$. This yields coefficients that adapt jointly to the fast-thinking update and its reasoning-aware perturbation weight.

Finally, we construct the merged model by applying the resulting coefficients to the fast-thinking task vector:
\begin{equation}
\theta^\star
=
\theta_S
+
\sum_b
\alpha_b^\star \Delta_{F,b},
\end{equation}
where $b$ spans all merged attention and FFN parameter groups; all remaining parameters are kept at their values in $\theta_S$. Attention coefficients follow from the allocation above. As FFNs lack head structure, we assign each layer the mean $Q/O$ coefficient, $\alpha_{\ell}^{\star,\mathrm{FFN}}=H^{-1}\sum_{h=1}^{H}\alpha_{\ell,h}^{\star,QO}$, where $H$ is the number of query heads per layer. Because FFNs transform the full hidden representation of each token rather than a head-specific subspace, their updates may affect the reasoning-to-rating mapping more broadly, motivating more conservative merging. In particular, validation shows that merging the final several FFN layers degrades the accuracy–efficiency trade-off, indicating that these layers are especially sensitive to changes in the reasoning-to-rating mapping. We therefore exclude layers 30–35 from FFN merging by setting their coefficients to zero in our experiments; results are reported in \S\ref{sec:rq3-ffn}.

\section{Experiments}
\newcommand{\std}[1]{\,{\tiny #1}}
\begin{table*}[t]
\centering
\caption{Main results across three datasets. Lower is better ($\downarrow$). Among merging methods, bold and underlined values denote the best and second-best results, respectively. MAE and Tok report 95\% $t$-based confidence intervals ($\pm$), while RMSE reports 95\% percentile-bootstrap confidence intervals ([\,]).}
\label{tab:main}
\footnotesize 
\setlength{\tabcolsep}{2.5pt}
\begin{tabular}{l ccc ccc ccc}
\toprule
& \multicolumn{3}{c}{\textbf{Amazon Book}}
& \multicolumn{3}{c}{\textbf{Yelp}}
& \multicolumn{3}{c}{\textbf{Amazon Music}} \\
\cmidrule(lr){2-4} \cmidrule(lr){5-7} \cmidrule(lr){8-10}
\textbf{Method}
& MAE$\downarrow$ & RMSE$\downarrow$ & Tok$\downarrow$
& MAE$\downarrow$ & RMSE$\downarrow$ & Tok$\downarrow$
& MAE$\downarrow$ & RMSE$\downarrow$ & Tok$\downarrow$ \\
\midrule
Qwen2.5-3B-Instruct
    & 0.7504$\pm$0.0160 & 0.9521\std{[.930,.974]} & 342.68$\pm$1.40
    & 0.9160$\pm$0.0230 & 1.1724\std{[1.142,1.203]} & 322.71$\pm$1.70
    & 0.6875$\pm$0.0290 & 0.8662\std{[.824,.908]} & 334.69$\pm$2.50 \\
RecZero ($\theta_S$)
    & 0.6650$\pm$0.0180 & 0.9331\std{[.906,.959]} & 313.58$\pm$1.25
    & 0.7769$\pm$0.0250 & 1.1054\std{[1.069,1.141]} & 314.54$\pm$1.30
    & 0.5433$\pm$0.0360 & 0.8510\std{[.796,.905]} & 344.28$\pm$2.25 \\
TALLRec ($\theta_F$)\footnotemark[1]
    & 0.6994$\pm$0.0255 & 1.1553\std{[1.122,1.187]} & 4.0
    & 0.8082$\pm$0.0340 & 1.3509\std{[1.307,1.394]} & 4.0
    & 0.7871$\pm$0.0310 & 0.9669\std{[.933,1.002]} & 4.0 \\
\midrule
\rowcolor{BlockPurple}
\multicolumn{10}{c}{\textit{Arithmetic and pruning-based merging}} \\
Average Merging
    & 0.6880$\pm$0.0170 & 0.9202\std{[.898,.941]} & 270.05$\pm$1.15
    & 0.7907$\pm$0.0235 & 1.0866\std{[1.053,1.120]} & 271.96$\pm$1.45
    & 0.5943$\pm$0.0325 & 0.8344\std{[.786,.882]} & 280.48$\pm$2.20 \\
Task Arithmetic
    & 0.6742$\pm$0.0180 & 0.9333\std{[.911,.956]} & 253.13$\pm$1.58
    & \underline{0.7723$\pm$0.0240} & 1.0872\std{[1.054,1.121]} & \textbf{256.11$\pm$2.40}
    & 0.5605$\pm$0.0335 & \underline{0.8265\std{[.779,.873]}} & \underline{269.44$\pm$2.55} \\
DARE
    & 0.7030$\pm$0.0170 & 0.9320\std{[.910,.954]} & 272.32$\pm$1.20
    & 0.8025$\pm$0.0235 & 1.0968\std{[1.063,1.130]} & 275.27$\pm$2.45
    & 0.6122$\pm$0.0315 & 0.8409\std{[.794,.888]} & 277.55$\pm$2.75 \\
DARE + TA
    & 0.6835$\pm$0.0175 & 0.9343\std{[.912,.956]} & \underline{252.81$\pm$1.53}
    & 0.7809$\pm$0.0240 & 1.0921\std{[1.058,1.125]} & 265.79$\pm$4.30
    & 0.5869$\pm$0.0340 & 0.8486\std{[.799,.898]} & \textbf{266.60$\pm$2.70} \\
DARE + TIES\footnotemark[2]
    & 0.7161$\pm$0.0210 & 0.9992\std{[.976,1.023]} & 261.39$\pm$2.20
    & 0.7591$\pm$0.0270 & 1.0855\std{[1.051,1.121]} & 266.21$\pm$2.20
    & 0.5947$\pm$0.0355 & 0.8651\std{[.822,.909]} & 257.75$\pm$3.25 \\
\midrule
\rowcolor{BlockPurple}
\multicolumn{10}{c}{\textit{Data-driven merging}} \\
AIM + TA
    & 0.6805$\pm$0.0175 & 0.9272\std{[.905,.949]} & 260.46$\pm$1.15
    & 0.7735$\pm$0.0235 & \underline{1.0788\std{[1.046,1.112]}} & 261.62$\pm$1.35
    & 0.5678$\pm$0.0330 & \textbf{0.8227\std{[.775,.871]}} & 280.71$\pm$2.20 \\
ACM + TA
    & 0.6932$\pm$0.0165 & \underline{0.9162\std{[.894,.938]}} & 282.74$\pm$1.15
    & 0.7987$\pm$0.0240 & 1.0995\std{[1.066,1.132]} & 290.03$\pm$1.65
    & 0.5871$\pm$0.0325 & 0.8333\std{[.786,.881]} & 304.72$\pm$2.35 \\
RAIN-Merging
    & \underline{0.6632$\pm$0.0180} & 0.9325\std{[.906,.959]} & 290.30$\pm$1.20
    & 0.7731$\pm$0.0250 & 1.1065\std{[1.070,1.143]} & 289.17$\pm$1.45
    & \underline{0.5424$\pm$0.0365} & 0.8527\std{[.799,.905]} & 315.63$\pm$2.25 \\
\midrule
\textbf{REAM (ours)}
    & \textbf{0.6338$\pm$0.0181} & \textbf{0.9116\std{[.888,.936]}} & \textbf{237.33$\pm$1.24}
    & \textbf{0.7564$\pm$0.0237} & \textbf{1.0649\std{[1.033,1.098]}} & \underline{258.19$\pm$1.25}
    & \textbf{0.5348$\pm$0.0351} & 0.8324\std{[.783,.882]} & 271.49$\pm$2.49 \\
Reduction vs.\ $\theta_S$
    & \textit{4.7\%} & \textit{2.3\%} & \textit{24.3\%}
    & \textit{2.6\%} & \textit{3.7\%} & \textit{17.9\%}
    & \textit{1.6\%} & \textit{2.2\%} & \textit{21.1\%} \\
\bottomrule
\end{tabular}
\end{table*}
\footnotetext[1]{TALLRec outputs a short numerical rating response; token count is not schema-comparable to other methods.}
\footnotetext[2]{DARE+TIES metrics are computed over successfully parsed outputs only and are not directly comparable.}

We organise our experiments around the following research questions (RQs), addressed in \S\ref{sec:rq1} to \S\ref{sec:rq3}, respectively:
\begin{itemize}[leftmargin=1.05em, labelsep=0.35em, itemsep=0pt, topsep=1pt]
    \item \textbf{RQ1:} Can general-purpose model merging shorten slow-thinking reasoning traces while preserving recommendation accuracy?
    \item \textbf{RQ2:} Does REAM achieve a better accuracy--efficiency trade-off than existing training-free merging baselines, including both data-driven and arithmetic-based methods?
    \item \textbf{RQ3:} How do REAM's individual design choices contribute to the resulting accuracy--efficiency trade-off?
\end{itemize}

\subsection{Experimental Setup}
\paragraph{Models and datasets.}
We evaluate on Amazon Book, Amazon Music, and Yelp using the Reason4Rec rating-prediction splits~\cite{fang2025reason4reclargelanguagemodels}. Each instance provides a user's historical ratings and reviews and a target item, with a 1--5 rating as the label. The slow-thinking model $\theta_S$ is RecZero~\cite{kong2025thinkrecommendationautonomousreasoningenhanced}, fine-tuned from Qwen2.5-3B-Instruct for structured reasoning. The fast-thinking model $\theta_F$ is an adapted TALLRec model~\cite{Bao_2023}, fine-tuned from the base Qwen2.5-3B-non-instruct to obtain a behaviourally distinct direct-prediction model, as an instruct-initialised model may retain substantial reasoning and instruction-following behaviour even when fine-tuned for direct rating prediction. (Appendix~\ref{app:tallrec_adaptation}). We use Qwen2.5-3B-Instruct as the shared base $\theta_B$ for both task vectors. 

\paragraph{Baselines.}
We compare REAM with the two source models and eight representative training-free baselines spanning arithmetic-based, pruning-based, and data-driven methods (Appendices~\ref{app:baselines} and~\ref{app:baseline_hyperparams}). Among the data-driven baselines, AIM and ACM derive one coefficient per layer from activation statistics, whereas RAIN-Merging operates at head granularity but bases its coefficients on instruction relevance rather than recommendation reasoning---the distinction REAM's allocation is built around. All baselines merge the same $\theta_S$ and $\theta_F$ using their recommended configurations.

\paragraph{Metrics.}
We report MAE and RMSE for rating accuracy and mean generated tokens per response (Tok) as a proxy for reasoning verbosity; lower is better for all metrics. Metrics are computed on parseable outputs (Appendix~\ref{app:experimental_configuration}); parse rates are near 100\% for all methods except DARE+TIES, as noted in Table~\ref{tab:main}.

\paragraph{Implementation details.}
REAM and all baselines share a common configuration (calibration, hyperparameters, hardware; Appendix~\ref{app:experimental_configuration}), with baseline-specific settings in Appendix~\ref{app:baseline_hyperparams} and REAM's computational overhead reported in Appendix~\ref{app:overhead}.

\subsection{Limits of General-Purpose Model Merging}
\label{sec:rq1}
General-purpose model merging provides a training-free means of transferring concise generation from a fast-thinking recommender to a slow-thinking one, but the accuracy--efficiency trade-off remains unclear. We therefore evaluate arithmetic- and pruning-based methods alongside data-driven merging approaches to determine how substantially they shorten reasoning traces and whether these efficiency gains consistently preserve recommendation accuracy.

\textbf{(1) General-purpose merging shortens reasoning, but does not reliably preserve accuracy.} On Book, Task Arithmetic and DARE+TA achieve substantial compression, reducing generation length to 253.13 and 252.81 tokens, respectively, but both worsen MAE relative to $\theta_S$ (0.6742 and 0.6835 vs.\ 0.6650). Average Merging and DARE exhibit the same trade-off, although with more modest compression (Table~\ref{tab:main}). Importantly, this pattern is not confined to arithmetic- or pruning-based methods: AIM+TA and ACM+TA also worsen MAE to 0.6805 and 0.6932, with ACM+TA underperforming most arithmetic baselines despite incorporating activation statistics. These results show that general-purpose merging can transfer concise generation, but does not consistently preserve the prediction behaviour of the slow-thinking model.

\textbf{(2) General-purpose merging can alter the distinctive rating behaviour of the slow-thinking model.} On Music, most general-purpose merging methods improve RMSE relative to $\theta_S$ while worsening MAE. Task Arithmetic, DARE, and DARE+TA achieve RMSE 0.8265--0.8486, compared with 0.8510 for $\theta_S$, but increase MAE from 0.5433 to 0.5605--0.6122. AIM+TA and ACM+TA exhibit the same divergence, with AIM+TA attaining the lowest RMSE among baselines (0.8227) while worsening MAE to 0.5678 (Table~\ref{tab:main}). Because RMSE penalises large errors more heavily than MAE, a lower RMSE alongside a higher MAE is consistent with a redistribution of errors---fewer extreme deviations but larger absolute errors on average---rather than an overall improvement in accuracy~\cite{Willmott2005AdvantagesOT}. This redistribution is consequential for slow-thinking recommenders, whose predictive advantage depends on distinguishing fine-grained user--item compatibility and grounding each rating in that assessment~\cite{fang2025reason4reclargelanguagemodels,kong2025thinkrecommendationautonomousreasoningenhanced}. Such a shift does not necessarily preserve this instance-specific judgement, suggesting that the merged model may alter the rating behaviour of $\theta_S$ across individual cases.


\subsection{Accuracy–Efficiency Trade-off of REAM}
\label{sec:rq2}
\begin{figure}[t]
    \centering
    \includegraphics[width=\columnwidth]{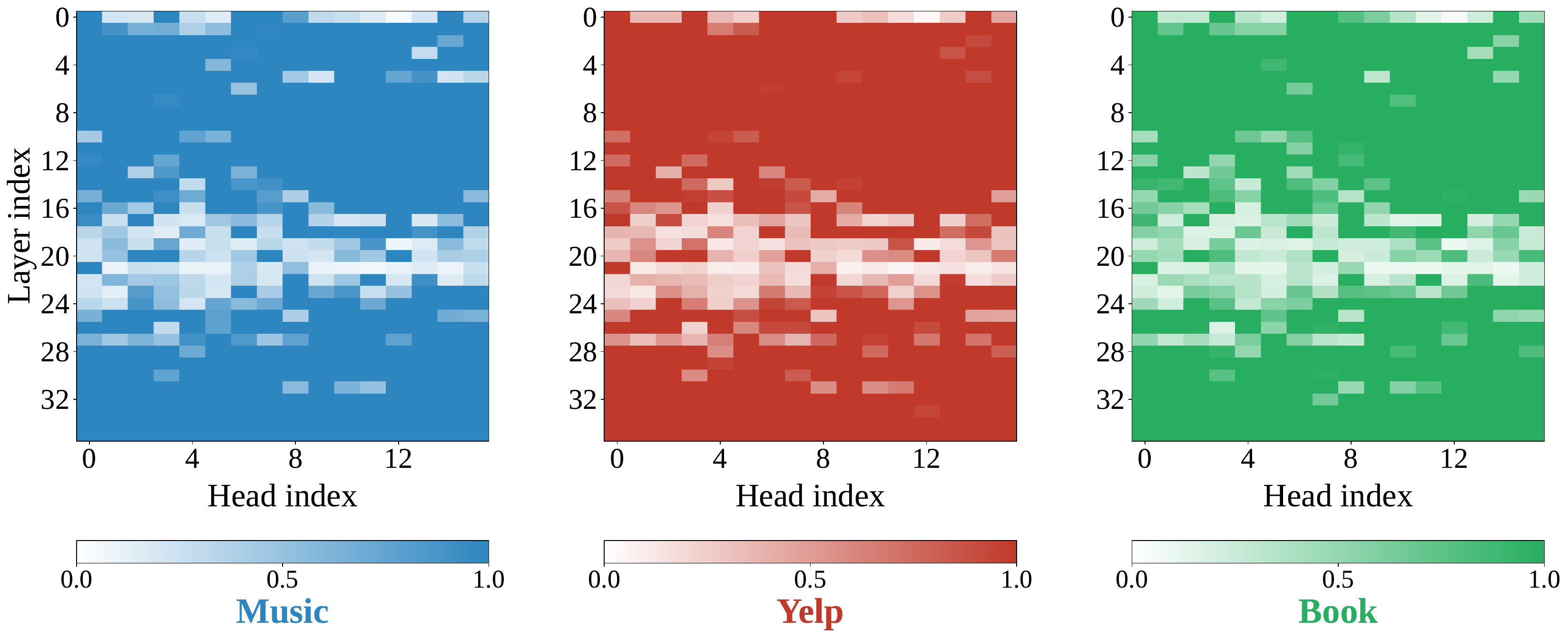}
    \caption{Per-head merge coefficients $\alpha^\star$ across layers and attention heads on Music, Yelp, and Book.}
    \label{fig:alpha_heatmap}
\end{figure}
Having established the need for more selective merging, we next evaluate whether REAM's reasoning-grounded, head-level allocation achieves a better accuracy--efficiency trade-off than existing model merging methods. Table~\ref{tab:main} compares REAM with the two source models and eight baselines across Book, Yelp, and Music.

\textbf{(1) REAM delivers the most consistent balance between accuracy and efficiency.} REAM is the only comparable merging method that maintains accuracy across MAE and RMSE while shortening reasoning on all three datasets. It achieves the best MAE and RMSE on Book and Yelp, the shortest trace on Book, and the best MAE on Music. Although several baselines attain stronger results on isolated metrics, none sustains these gains across both accuracy measures and reasoning length. Specifically, AIM+TA achieves a lower RMSE on Music (0.8227 vs.\ 0.8324), but worsens MAE relative to REAM (0.5678 vs.\ 0.5348) and performs less favourably on the other datasets. Taken together, these results establish REAM as the most consistent method across all three recommendation domains.

\textbf{(2) REAM's advantage reflects both reasoning-aware signals and head-level allocation.} For reasoning compression in recommender systems, effective coefficient design requires a signal grounded in recommendation reasoning and sufficient granularity to act on it. Concretely, AIM+TA and ACM+TA derive coefficients from activation statistics but assign a single value per layer, thereby applying the same merge strength to heads with substantially different reasoning roles. This coarse allocation yields a consistently weaker accuracy--efficiency trade-off: across all three datasets, AIM+TA and ACM+TA produce longer traces and higher MAE than REAM (e.g., 260.46 and 282.74 vs.\ 237.33 tokens, and 0.6805 and 0.6932 vs.\ 0.6338 MAE on Book). The observed head-level heterogeneity helps explain this gap: on Book, the top 5\% of retrieval-critical heads hold 78.0\% of retrieval mass (Gini$=0.922$), whereas decision faithfulness is less concentrated (Gini$=0.553$), indicating that reasoning importance is distributed unevenly across heads (Appendix~\ref{app:critical_heads_sparsity}). Accordingly, Figure~\ref{fig:alpha_heatmap} reveals substantial coefficient variation within individual layers. RAIN-Merging operates at head granularity, but derives its coefficients from instruction relevance rather than recommendation reasoning and underperforms REAM on every metric across all three datasets (Table~\ref{tab:main}). Together, these comparisons show that head-level control is most effective when paired with a reasoning-grounded allocation criterion.

\subsection{Impact of Key Components}
\label{sec:rq3}

\subsubsection{Component Ablation}
\label{sec:rq3-ablation}
We ablate REAM's three allocation signals---retrieval criticality $\kappa$, decision faithfulness, and Fisher update sensitivity $d$---on the full Music and Yelp test sets (Table~\ref{tab:component_ablation}). Removing Fisher sensitivity has the largest effect on both datasets, increasing MAE to 0.5533 on Music (vs.\ 0.5348) and 0.7616 on Yelp (vs.\ 0.7564). Removing faithfulness has a smaller effect on Music MAE (0.5350) and slightly improves MAE on Yelp (0.7544), but on both datasets increases RMSE and reasoning length. Together with the near-zero correlation between $\kappa$ and faithfulness (Spearman $|\rho|\leq0.018$ in every domain; Appendix~\ref{app:critical_heads_independence}), these results support complementary roles: Fisher sensitivity captures vulnerability to the update, while $\kappa$ and faithfulness identify distinct retrieval- and decision-related risks whose contributions are not fully reflected by any single metric.

\begin{table}[t]
\centering
\caption{Leave-one-out ablation on Amazon Music and Yelp. $\downarrow$ indicates lower is better. Bold marks best value per column.}
\label{tab:component_ablation}
\footnotesize
\begin{tabular*}{\columnwidth}{@{\extracolsep{\fill}}l ccc ccc@{}}
\toprule
& \multicolumn{3}{c}{\textbf{Amazon Music}} & \multicolumn{3}{c}{\textbf{Yelp}} \\
\cmidrule(lr){2-4} \cmidrule(lr){5-7}
\textbf{Config.} & MAE & RMSE & Tok & MAE & RMSE & Tok \\
\midrule
\textbf{Full} & \textbf{0.5348} & 0.8324 & \textbf{271.49} & 0.7564 & \textbf{1.0649} & \textbf{258.19} \\
\midrule
w/o $\kappa$ & 0.5499 & 0.8320 & 271.53 & 0.7599 & 1.0807 & 259.55 \\
w/o $\mathrm{faith}$ & 0.5350 & 0.8350 & 271.98 & \textbf{0.7544} & 1.0716 & 263.72 \\
w/o $d$ & 0.5533 & \textbf{0.8252} & 271.89 & 0.7616 & 1.0854 & 262.62 \\
\bottomrule
\end{tabular*}
\end{table}

\begin{figure}[t]
    \centering
    \includegraphics[width=\columnwidth]{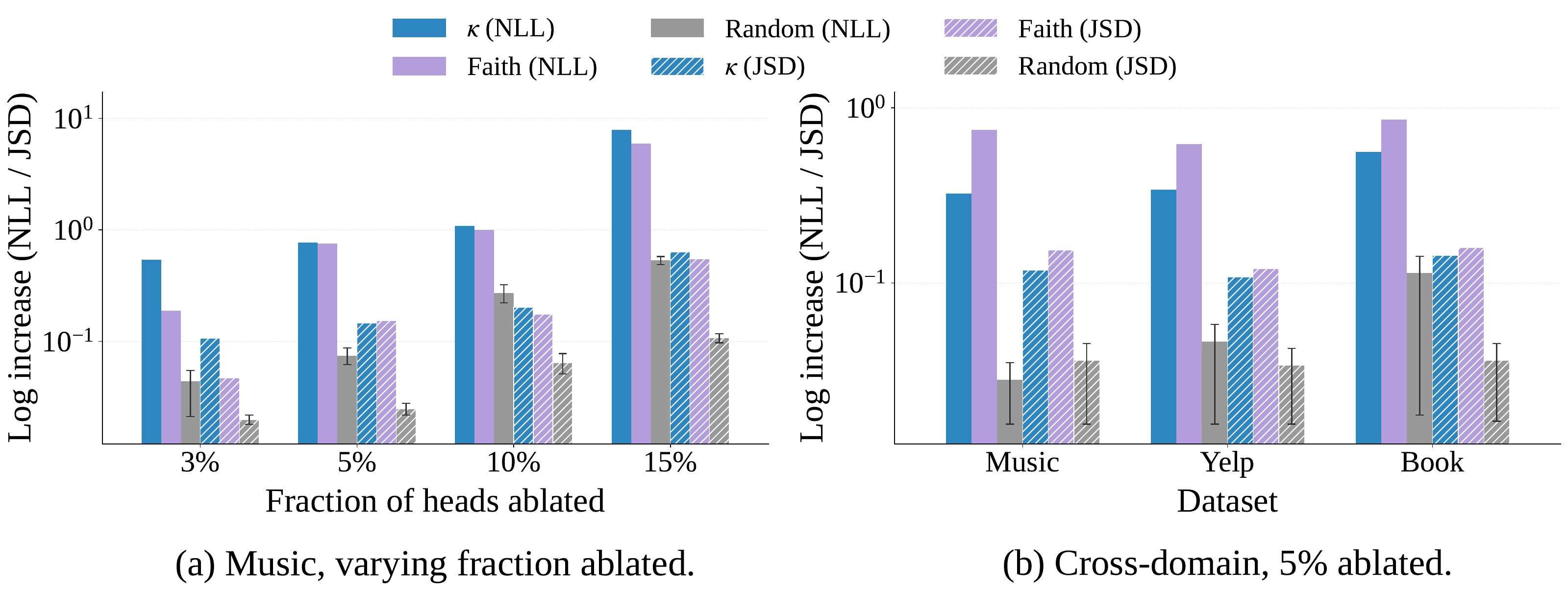}
\caption{NLL and JSD increase (log scale) from ablating retrieval-critical ($\kappa$) vs.\ random heads.}
    \label{fig:ablation_validation}
\end{figure}

\subsubsection{Causal Validation of Head-Level Signals}
\label{sec:rq3-validation}
We test whether the heads identified by $\kappa$ and decision faithfulness are causally important by mean-ablating the top 5\% under each signal and comparing held-out NLL and JSD with size-matched random-head ablations averaged over three draws. For NLL, the ratio compares increases from the unablated baseline; for JSD, it directly compares the divergence under critical- and random-head ablation (Appendix~\ref{app:critical_heads_causal}). Relative to random ablation, ablating $\kappa$-critical heads increases NLL/JSD by $11.57\times$/$3.29\times$ on Music, $4.93\times$/$3.96\times$ on Book, and $7.35\times$/$3.19\times$ on Yelp (Figure~\ref{fig:ablation_validation}). Faithfulness-critical heads cause similarly large increases of $9.65\times$/$5.64\times$, $11.16\times$/$5.20\times$, and $6.63\times$/$4.19\times$, respectively. Despite these comparable causal effects, the two signals' top-5\% head sets are nearly disjoint (Jaccard $\leq1.75\%$; Table~\ref{tab:independence_stats}), while each remains stable across datasets (mean cross-domain Jaccard $87.1\%$ for $\kappa$ and $79.4\%$ for faithfulness; Appendix~\ref{app:critical_heads_sparsity}). Together, these results show that the two signals identify complementary and transferable head-level roles rather than redundant rankings.

\begin{figure}[t]
    \centering
    \includegraphics[width=\columnwidth]{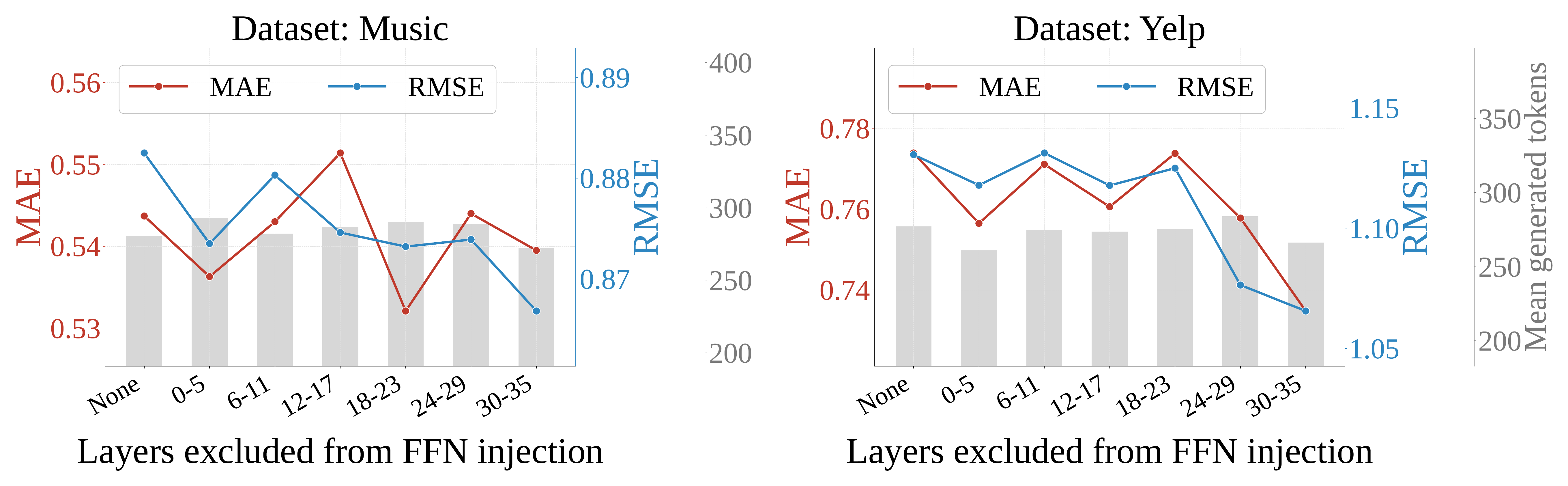}
    \caption{FFN exclusion sweep on Music and Yelp subsets. Bars show Tok; lines show MAE and RMSE.}
    \label{fig:ffn_exclusion}
\end{figure}

\subsubsection{FFN Exclusion}
\label{sec:rq3-ffn}
Because FFNs lack head structure, their merge coefficients are assigned at layer granularity. We assess whether the fast-thinking update can be safely applied across depth by sweeping six-layer exclusion ranges on the Music and Yelp validation subsets (Figure~\ref{fig:ffn_exclusion}). Excluding the final six layers, $\mathcal{L}_{\mathrm{excl}}=\{30,\ldots,35\}$, gives the most consistent accuracy--efficiency trade-off: best RMSE on both datasets, best MAE on Yelp, and shortest trace on Music. This suggests late FFNs are especially sensitive to the fast-thinking update, consistent with FFNs' direct role in shaping model outputs~\cite{geva2021transformerfeedforwardlayerskeyvalue}, where perturbation near the output may disrupt the reasoning-to-rating mapping. We therefore use layers 30--35 as the fixed FFN exclusion window for all Qwen2.5-3B evaluations (\S\ref{sec:allocation}). The broader late-layer exclusion strategy generalises across model scales and families (Appendix~\ref{app:ffn_generalization}).

\begin{figure}[t]
    \centering
    \includegraphics[width=\columnwidth]{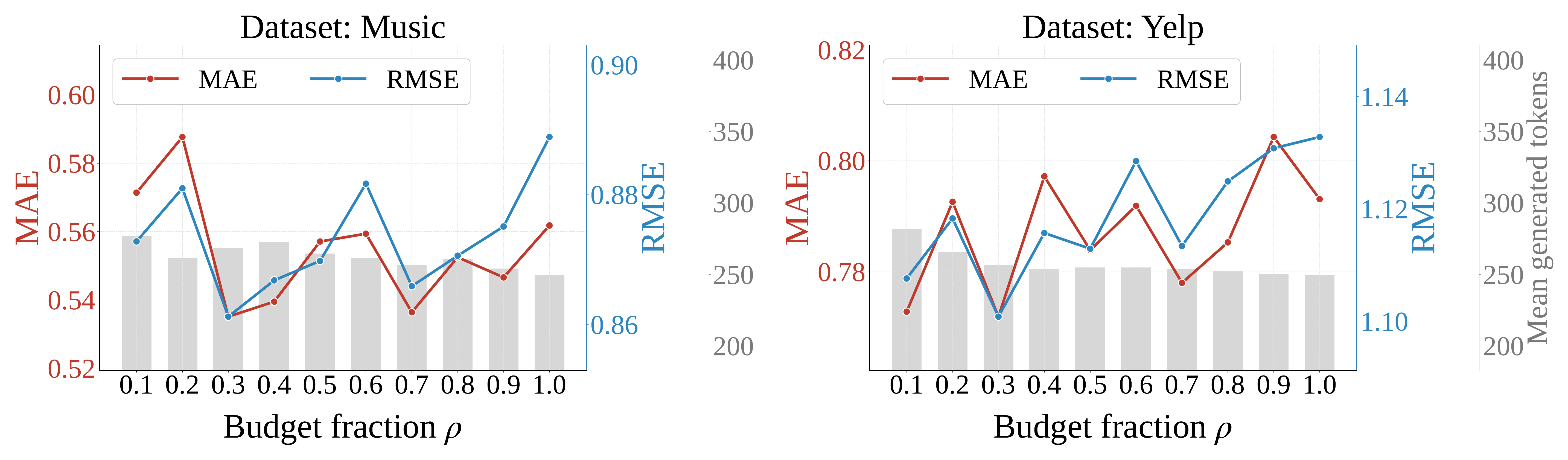}
    \caption{Hyperparameter analysis: MAE, RMSE, and Tok vs. budget
    fraction $\rho$, on Music and Yelp subsets.}
    \label{fig:rho_sweep}
\end{figure}

\subsubsection{Hyperparameter Sensitivity}
\label{sec:rq3-hyperparam}
We sweep the budget fraction $\rho$, which controls how much of the fast-thinking update the allocation can absorb, over $\rho \in [0.1,1.0]$ on the Music and Yelp validation subsets (Figure~\ref{fig:rho_sweep}). Performance is non-monotonic, confirming that $\rho$ is a true allocation budget: too little update limits trace shortening, while too much can perturb reasoning-sensitive components. Our fixed choice, $\rho=0.3$, consistently lies near the best accuracy--efficiency trade-off on both datasets.

\begin{table}[t]
\centering
\caption{Generalisability of REAM across model scales, base model families, and reasoning backbones (Yelp). $\downarrow$ indicates lower is better. CI notation follows Table~\ref{tab:main}.}
\label{tab:Generalisability}
\footnotesize
\renewcommand{\arraystretch}{1.05}
\begin{tabular*}{\columnwidth}{@{\extracolsep{\fill}}l ccc@{}}
\toprule
\textbf{Method} & MAE$\downarrow$ & RMSE$\downarrow$ & Tok$\downarrow$ \\
\midrule
\multicolumn{4}{l}{\textit{Qwen2.5-7B-Instruct}} \\[1pt]
RecZero ($\theta_S$)  & \textbf{0.7626}$\pm$0.024 & \textbf{1.0757}\std{[1.040,1.112]} & 336.02$\pm$1.40 \\
Task Arithmetic        & 0.8134$\pm$0.025 & 1.1327\std{[1.098,1.167]} & 302.37$\pm$1.95 \\
\textbf{REAM (ours)}   & 0.7910$\pm$0.024 & 1.0864\std{[1.053,1.120]} & \textbf{288.46}$\pm$1.35 \\
\midrule
\multicolumn{4}{l}{\textit{Llama-3.2-3B-Instruct}} \\[1pt]
RecZero ($\theta_S$)  & 0.8059$\pm$0.024 & 1.0977\std{[1.069,1.127]} & 323.79$\pm$1.85 \\
Task Arithmetic        & \textbf{0.7269}$\pm$0.030 & 1.1395\std{[1.101,1.178]} & \textbf{254.18}$\pm$1.05 \\
\textbf{REAM (ours)}   & 0.7769$\pm$0.026 & \textbf{1.0895}\std{[1.055,1.126]} & 257.18$\pm$2.20 \\
\midrule
\multicolumn{4}{l}{\textit{RecOne (Qwen2.5-3B-Instruct)}} \\[1pt]
RecOne ($\theta_S$)    & 0.7395$\pm$0.027 & 1.1310\std{[1.094,1.168]} & 365.69$\pm$0.65 \\
Task Arithmetic        & 0.6885$\pm$0.028 & 1.1200\std{[1.083,1.158]} & 321.68$\pm$1.15 \\
\textbf{REAM (ours)}   & \textbf{0.6876}$\pm$0.028 & \textbf{1.1137}\std{[1.077,1.151]} & \textbf{309.70}$\pm$1.10 \\
\bottomrule
\end{tabular*}
\end{table}

\subsubsection{Generalisability}
\label{sec:rq3-Generalisability}
REAM's transferability across model scale, base-model family, and training paradigm is evaluated using the same hyperparameters as Appendix~\ref{app:experimental_configuration}, without re-tuning (Table~\ref{tab:Generalisability}). On Qwen2.5-7B, REAM shortens $\theta_S$'s trace from 336.02 to 288.46 tokens and outperforms Task Arithmetic in MAE and RMSE, although it remains less accurate than $\theta_S$. On Llama-3.2-3B, REAM improves on $\theta_S$ across all three metrics, and holds a lower RMSE than Task Arithmetic despite Task Arithmetic's shorter trace and lower MAE---consistent with the same MAE--RMSE divergence observed in \S\ref{sec:rq1}, where aggressive compression can come at the cost of rating stability. On RecOne~\cite{kong2025thinkrecommendationautonomousreasoningenhanced}, which learns the same typed-segment format through SFT followed by RL rather than RecZero's pure-RL training, REAM outperforms both $\theta_S$ and Task Arithmetic on all three metrics. These results support REAM's transferability across model scales, families, and training paradigms, while leaving room for improvement in some settings.

\section{Conclusion}
This work shows that reasoning compression in slow-thinking recommender systems depends not only on how much of the fast-thinking update is merged, but also on which components receive it. To our knowledge, REAM is the first merging framework designed for this setting, assigning each head a merge coefficient according to its retrieval criticality, decision faithfulness, and sensitivity to parameter change. The two reasoning signals identify causally important yet largely disjoint head sets, supporting their complementary rather than redundant contributions. Across three recommendation datasets, REAM shortens reasoning while maintaining accuracy, achieving a more favourable accuracy--efficiency trade-off than the evaluated training-free merging baselines. The remaining accuracy gap at larger model scale highlights an important direction for extending selective merging to larger recommenders.

\bibliographystyle{ACM-Reference-Format}
\bibliography{references}

\appendix

\section{Implementation Details}
\subsection{Fast-Thinking Model Adaptation}
\label{app:tallrec_adaptation}

TALLRec~\cite{Bao_2023} was originally trained with LoRA~\cite{hu2021loralowrankadaptationlarge} for binary recommendation feedback: given a user's interaction history and a target item, the model predicts whether the user would like the item, e.g., \texttt{Output: Yes.}

We adapt TALLRec to serve as the fast-thinking model $\theta_F$ in two ways. First, we change the task from binary feedback to direct rating prediction. The adapted model outputs a single rating on a 1--5 scale without intermediate reasoning, making its output directly comparable to the final \textsc{rate} value produced by $\theta_S$. Appendix~\ref{app:format} shows this adapted output alongside RecZero's structured reasoning trace on the same instance.

Second, we fully fine-tune all parameters of TALLRec rather than using LoRA, since our merging approach requires task-vector slices expressed as direct differences over full weight matrices --- $W_Q$, $W_K$, $W_V$, $W_O$, and the FFN weights --- rather than the low-rank adapter factors LoRA produces.

\subsection{Prompt Format Examples}
\label{app:format}

Both models receive the same input --- a user's history of previously rated items and reviews, followed by the target item --- and differ only in how they produce a rating. Prompts below are abbreviated for illustration; the actual input includes the full user history and target item description.

\subsubsection{RecZero ($\theta_S$): Slow-Thinking Structured Reasoning}

\begin{mdframed}[linewidth=0.8pt, innerleftmargin=10pt, innerrightmargin=10pt, innertopmargin=8pt, innerbottommargin=8pt, backgroundcolor=gray!5]
{\small\ttfamily\raggedright\sloppy
\setlength{\parindent}{0pt}
\textbf{Input:} User history with previously rated books and reviews, followed
by the target item:
\smallskip

Target item: Shifting Fates (Urban Fantasy Shifter Romance); Rating: \textbf{?}
\medskip

\textbf{Output:}
\smallskip

\texttt{<analyze\_user>} The user prefers emotionally engaging romance, morally
complex characters, strong world-building, and unconventional romantic settings.
\texttt{</analyze\_user>}
\smallskip

\texttt{<analyze\_item>} The target item is an urban fantasy shifter romance with
a protective female lead and survival-focused premise.

\texttt{</analyze\_item>}
\smallskip

\texttt{<match>} The target item matches the user's preference for strong romantic
tension, fantasy settings, and distinctive world-building, although the multi-author
format may introduce some uncertainty. 

\texttt{</match>}
\smallskip

\texttt{<rate>} 4.0 \texttt{</rate>}
}
\medskip

\noindent\textit{\small RecZero generates a four-segment output containing a three-segment reasoning trace before
producing the final rating.}
\end{mdframed}

\subsubsection{TALLRec ($\theta_F$): Fast-Thinking Direct Prediction}

\begin{mdframed}[linewidth=0.8pt, innerleftmargin=10pt, innerrightmargin=10pt, innertopmargin=8pt, innerbottommargin=8pt, backgroundcolor=gray!5]
{\small\ttfamily\raggedright\sloppy
\setlength{\parindent}{0pt}
\textbf{Instruction:} Predict rating for the last item.
\medskip

\textbf{Input:} Same user history and target item as above.
\medskip

\textbf{Output:} 4.0/5.0
}
\medskip

\noindent\textit{\small TALLRec predicts the rating directly without generating intermediate reasoning, producing only a short numerical response.}
\end{mdframed}

\subsection{Experimental Configuration}
\label{app:experimental_configuration}
For all merging methods, task vectors are computed relative to the shared base model $\theta_B$. Coefficient-based methods are evaluated using the common candidate range reported in Appendix~\ref{app:baseline_hyperparams}. REAM's reasoning-importance and Fisher-sensitivity statistics are estimated separately for each dataset using a calibration set of $|\mathcal{D}_{\mathrm{slow}}|=500$ examples. We fix $\bar{\alpha}=1$, $\gamma=3$, $\rho=0.3$, and $\delta=10^{-8}$ across all datasets. All experiments are conducted on a single NVIDIA A40 GPU.

\paragraph{Parseability criteria.} For models with typed reasoning, an output is parseable if it contains all required segments and a valid \texttt{<rate>...</rate>} span with an extractable rating; for TALLRec, only an extractable numeric rating is required.

\subsection{Computational Overhead of REAM}
\label{app:overhead}

Table~\ref{tab:overhead} reports the cost of REAM's full pipeline --- retrieval-criticality and decision-faithfulness estimation, Fisher-sensitivity computation, and coefficient allocation --- measured on a single NVIDIA A40 GPU, the same hardware used throughout our experiments (Appendix~\ref{app:experimental_configuration}). The full pipeline completes in approximately 41 minutes (0.69 GPU-hours) per dataset. Fisher-sensitivity estimation and the Q/O and K/V aggregation dominate this cost, accounting for 73.4\% of the total, while retrieval-criticality and faithfulness estimation account for 25.2\%, and the constrained-optimization allocation step is negligible at 1.4\%.

This cost is substantially lower than training the two source models. RecZero ($\theta_S$) requires 4.7 hours of RL training on 2 A40 GPUs (9.4 GPU-hours), and TALLRec ($\theta_F$) requires 6.3 hours of supervised fine-tuning on 1 A40 GPU (6.3 GPU-hours), for a combined 15.7 GPU-hours. REAM's entire pipeline therefore costs approximately 4\% of this combined training budget, and---unlike training a new slow-thinking recommender---is incurred only once per dataset, requires no gradient updates to either source model, and involves no changes to decoding.

\begin{table}[h]
\centering
\caption{Computational cost of REAM's training-free pipeline vs.\ fine-tuning the slow- and fast-thinking source models from their respective pretrained checkpoints. GPU-hours account for the number of GPUs used per training run.}
\label{tab:overhead}
\small
\begin{tabular}{lccc}
\toprule
Stage & Wall-clock time & GPUs & GPU-hours \\
\midrule
Reasoning-aware ($R_{\ell,h}$ + faith) & 10.4 min & 1 & 0.17 \\
Update sensitivity & 30.2 min & 1 & 0.50 \\
Allocation + merge & 0.6 min & 1 & 0.01 \\
\midrule
\textbf{REAM total} & \textbf{41.2 min} & 1 & \textbf{0.69} \\
\midrule
RecZero ($\theta_S$) training & 4.7 h & 2 & 9.4 \\
TALLRec ($\theta_F$) training & 6.3 h & 1 & 6.3 \\
\textbf{Combined training cost} & --- & --- & \textbf{15.7} \\
\bottomrule
\end{tabular}
\end{table}

\subsection{Baseline Descriptions}
\label{app:baselines}

Average Merging, Task Arithmetic (TA)~\cite{ilharco2023editingmodelstaskarithmetic}, DARE~\cite{yu2024languagemodelssupermario}, DARE+TA, and DARE+TIES~\cite{yadav2023tiesmergingresolvinginterferencemerging} are implemented using the Long-to-Short codebase~\cite{wu2025unlockingefficientlongtoshortllm}\footnote{\url{https://github.com/hahahawu/Long-to-Short-via-Model-Merging}}; AIM~\cite{nobari2025activationinformedmerginglargelanguage}, ACM~\cite{yao2025activationguidedconsensusmerginglarge}, and RAIN-Merging~\cite{huang2026rainmerginggradientfreemethodenhance} use their official codebases.\footnote{AIM: \url{https://github.com/ahnobari/ActivationInformedMerging}; ACM: \url{https://github.com/starrYYxuan/ACM}; RAIN-Merging: \url{https://github.com/K1nght/RAIN-Merging}.}

\textbf{Average Merging} directly averages $\theta_S$ and $\theta_F$'s parameters. \textbf{TA} adds task vectors (parameter differences from a shared base) with a uniform coefficient. \textbf{DARE} randomly drops and rescales task-vector entries before merging; we evaluate it alone, combined with TA (DARE+TA), and combined with TIES (DARE+TIES). \textbf{TIES} trims small entries and resolves sign conflicts across task vectors before merging; TIES's own trimming fraction is denoted $k_{\mathrm{TIES}}$. TIES alone consistently failed to produce usable output in our preliminary testing --- the same sign-conflict instability likely explains DARE+TIES's lower parse rate noted in Table~\ref{tab:main} --- so we report only DARE+TIES as the representative TIES-based baseline.

\textbf{AIM} sets layer-wise coefficients from base-model activation magnitudes; \textbf{ACM} instead uses mutual information between base and fine-tuned activations. Both are evaluated with TA. \textbf{RAIN-Merging}, the closest head-level baseline, projects updates onto the null space of reasoning-token activations and scales head coefficients by instruction-attention alignment --- grounded in instruction relevance rather than each head's role in evidence retrieval and decision grounding.

\subsection{Baseline Hyperparameters}
\label{app:baseline_hyperparams}

For each baseline, we use the hyperparameters recommended in its original release.

\begin{table}[h]
\centering
\caption{Hyperparameters used for each baseline, taken from
the original method's recommended configuration.}
\small
\begin{tabular}{ll}
\toprule
Method & Hyperparameters \\
\midrule
Average Merging & --- \\
Task Arithmetic & $\alpha = 0.7$ \\
DARE & $p = 0.3$ \\
DARE + TA & $p = 0.3$, $\alpha = 0.7$ \\
DARE + TIES & $p = 0.3$, $k_{\mathrm{TIES}} = 0.8$, $\alpha = 1.0$ \\
AIM + TA & $\omega = 0.4$ \\
ACM + TA & $t = 0.7$ \\
RAIN-Merging & 1000 samples, last 27 layers, ridge $=10^{-4}$ \\
\bottomrule
\end{tabular}
\end{table}

\begin{figure}[t]
    \centering
    \includegraphics[width=\columnwidth]{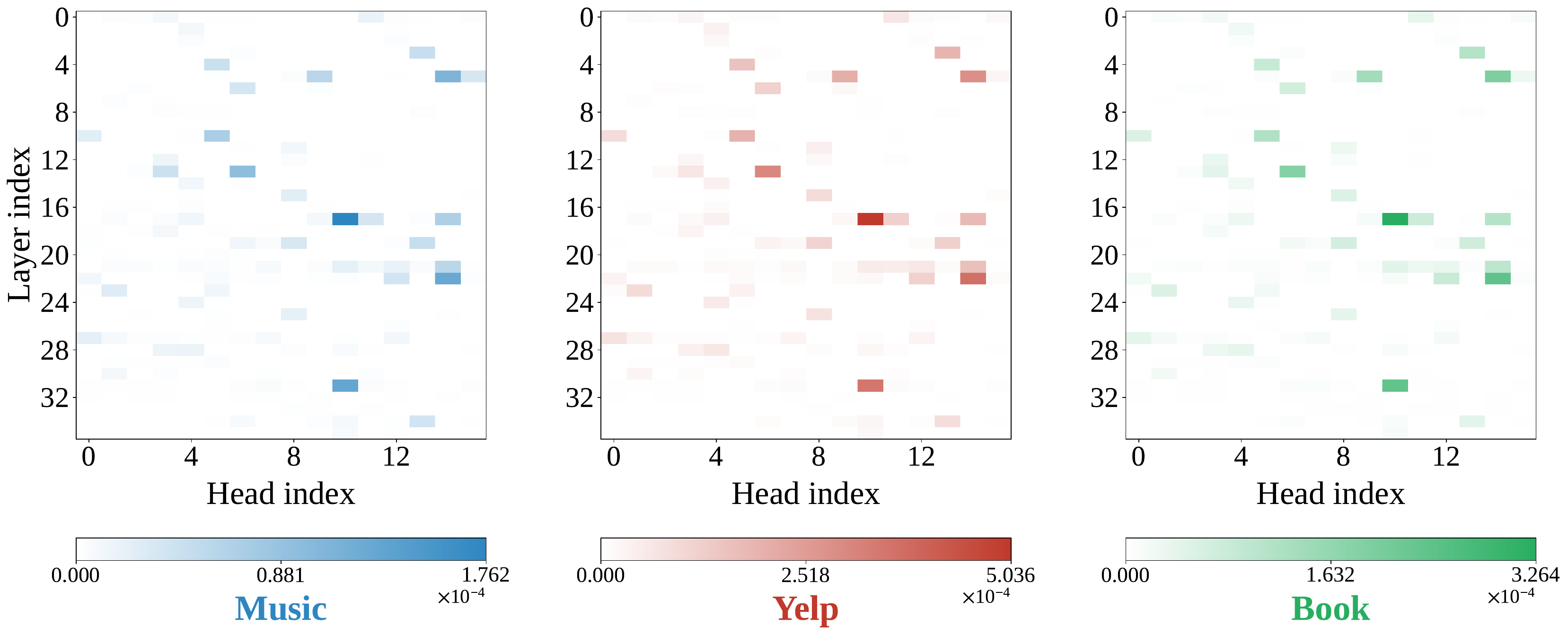}
    \caption{Retrieval criticality $R_{\ell,h}$ per head, across layers, on Music, Yelp, and Book.}
    \label{fig:retrieval_heatmap}
\end{figure}

\begin{figure}[t]
    \centering
    \includegraphics[width=\columnwidth]{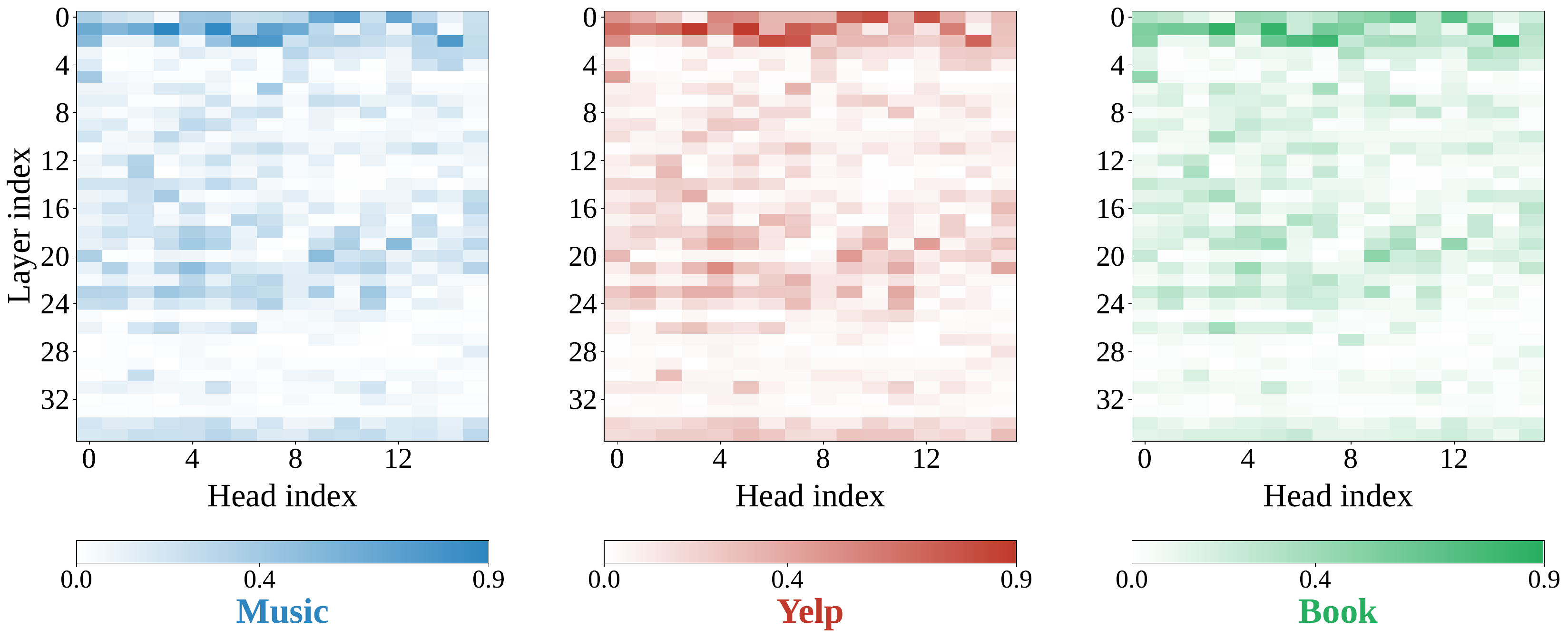}
    \caption{Decision faithfulness $\mathrm{faith}_{\ell,h}$ per head, across layers, on Music, Yelp, and Book.}
    \label{fig:faithfulness_heatmap}
\end{figure}

\subsection{Applicability to Other Reasoning Templates}
\label{app:template_generality}

REAM does not depend on RecZero's specific segment names or four-stage schema. For another structured template, its head-level analysis can be applied to identifiable regions containing the information required at each reasoning stage, including retrieved, collaborative, or multimodal context. The region connecting the preceding reasoning to the final prediction can likewise be identified through perturbation analysis. We evaluate REAM only under RecZero's schema; fully unstructured traces would require an additional segmentation procedure, which we leave to future work.

\subsection{Calibration Set Size and Estimate Stability}
\label{app:calibration_stability}

To assess whether $|\mathcal{D}_{\mathrm{slow}}|=500$ calibration samples are sufficient for stable retrieval-criticality estimates, we conduct two complementary analyses on Music.

\begin{figure}[h]
    \centering
    \includegraphics[width=0.75\columnwidth]{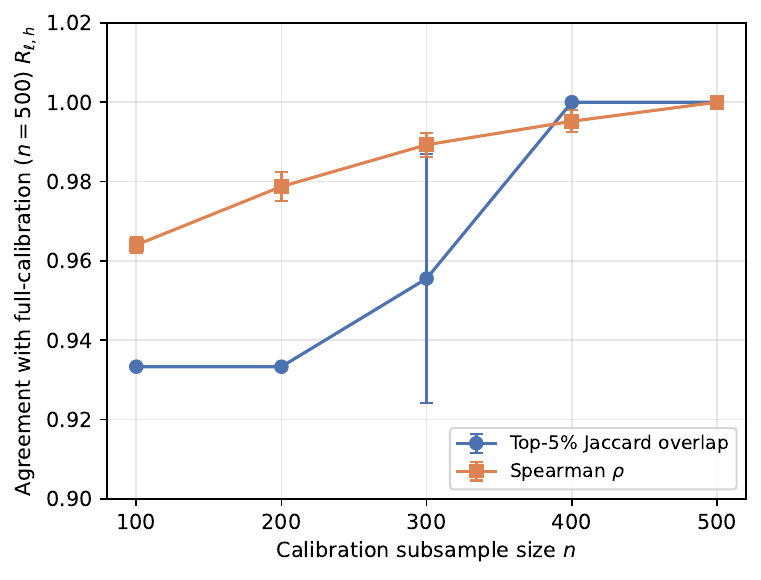}
    \caption{Agreement between subsample and full-set estimates of $R_{\ell,h}$ across calibration sizes on Music.}
    \label{fig:calibration_stability}
\end{figure}

\paragraph{Subsample stability.} We recompute $R_{\ell,h}$ using random subsamples of 100, 200, 300, and 400 examples, with three seeds per size, and compare their top-5\% critical-head sets and full rankings with those obtained from the full 500-sample set. Table~\ref{tab:subsample_stability} and Figure~\ref{fig:calibration_stability} report the mean Jaccard overlap and Spearman rank correlation. Both increase with sample size: at $n=300$, the top-head overlap reaches 0.956 and the rank correlation reaches 0.989, while $n=400$ recovers the full top-5\% set and achieves a correlation of 0.995.

\begin{table}[h]
\centering
\caption{Agreement between subsample and full-set estimates of $R_{\ell,h}$ on Music, averaged over three seeds.}
\label{tab:subsample_stability}
\small
\begin{tabular}{lcc}
\toprule
Subsample size & Mean top-5\% Jaccard & Mean Spearman $\rho$ \\
\midrule
100 & 0.933 & 0.964 \\
200 & 0.933 & 0.979 \\
300 & 0.956 & 0.989 \\
400 & 1.000 & 0.995 \\
500 (full) & 1.000 & 1.000 \\
\bottomrule
\end{tabular}
\end{table}

\paragraph{Bootstrap stability.} We further draw 1{,}000 bootstrap samples with replacement from the full 500-example calibration set and recompute $R_{\ell,h}$ for each draw. The top-5\% critical heads (29 of 576) have a median relative confidence-interval width of 14.95\%, compared with 20.88\% across all heads. Moreover, 96.6\% of these heads reappear in the bootstrap top-5\% set in at least 900 of the 1{,}000 draws.

Together, these analyses show that retrieval-criticality estimates stabilise before 500 samples and are most reliable for the heads REAM is designed to protect.

\subsection{Retrieval Criticality and Decision Faithfulness: Sparsity and Causal Validation}
\label{app:critical_heads}
We analyse $R_{\ell,h}$ and $\mathrm{faith}_{\ell,h}$ (\S\ref{sec:head_risk}) on $\mathcal{D}_{\mathrm{slow}}$, comprising 500 validation traces with correct ratings and all reasoning segments. For $R_{\ell,h}$, the evidence region $K^{(s)}(x)$ is defined once per trace and shared across generation steps.

\subsubsection{Sparsity and Cross-Domain Stability}
\label{app:critical_heads_sparsity}
Retrieval criticality is highly concentrated yet distributed across model depth (Figure~\ref{fig:retrieval_heatmap}), while decision faithfulness exhibits a different spatial pattern (Figure~\ref{fig:faithfulness_heatmap}): its mass is less concentrated but the top-5\% heads are more localised within early-to-middle layers. Table~\ref{tab:sparsity_stats} summarizes both signals' concentration and layer span across the three datasets. On Book, the top 5\% of heads (29 of 576) account for 78.0\% of retrieval mass, substantially exceeding both uniform allocation (Gini $=0$) and a random-noise null (i.i.d.\ $U(0,1)$, expected Gini $\approx 1/3$). Both signals' top-5\% head sets are stable across datasets: mean cross-domain Jaccard is $0.871$ for $R_{\ell,h}$ and $0.794$ for $\mathrm{faith}_{\ell,h}$.

\begin{table}[h]
\centering
\caption{Sparsity of the two head-level signals across datasets. Top-5\% mass is the share of total retrieval (or faithfulness) mass held by the top 5\% of heads.}
\label{tab:sparsity_stats}
\small
\begin{tabular}{llccc}
\toprule
Signal & Domain & Top-5\% mass & Gini & Layer span \\
\midrule
\multirow{3}{*}{$R_{\ell,h}$ (retrieval)} & Music & 77.6\% & 0.920 & 3--34 \\
 & Book & 78.0\% & 0.922 & 3--34 \\
 & Yelp & 77.1\% & 0.920 & 3--34 \\
\midrule
\multirow{3}{*}{$\mathrm{faith}_{\ell,h}$ (faithfulness)} & Music & 22.6\% & 0.548 & 0--23 \\
 & Book & 23.6\% & 0.553 & 0--23 \\
 & Yelp & 24.0\% & 0.556 & 0--23 \\
\bottomrule
\end{tabular}
\end{table}

\subsubsection{Independence Between Signals}
\label{app:critical_heads_independence}
Table~\ref{tab:independence_stats} reports the Spearman correlation and top-5\% head-set overlap between $\kappa_{\ell,h}$ and $\mathrm{faith}_{\ell,h}$ in each domain. Correlation is negligible everywhere, and the top-5\% head sets are correspondingly near-disjoint --- consistent with the layer-distribution contrast in Table~\ref{tab:sparsity_stats}: the two signals capture distinct, largely independent reasoning roles.

\begin{table}[h]
\centering
\caption{Independence between $\kappa_{\ell,h}$ and $\mathrm{faith}_{\ell,h}$: Spearman correlation and top-5\% head-set overlap.}
\label{tab:independence_stats}
\small
\begin{tabular}{lccc}
\toprule
Domain & Spearman $\rho$ & $p$-value & Top-5\% Jaccard \\
\midrule
Music & 0.018 & 0.667 & 0.0\% (0 shared) \\
Book & $-0.001$ & 0.976 & 0.0\% (0 shared) \\
Yelp & $-0.012$ & 0.769 & 1.75\% (1 shared) \\
\bottomrule
\end{tabular}
\end{table}

\subsubsection{Causal Validation}
\label{app:critical_heads_causal}
To confirm these heads are causally, not merely correlationally, critical, we mean-ablate them. For head $(\ell,h)$, let $o_{\ell,h}(x,t)\in\mathbb{R}^{d_h}$ denote its output slice at output-position token $t$ of sample $x$. We precompute a reference mean activation over $\mathcal{D}_\text{slow}$,
\begin{equation}
    \bar{o}_{\ell,h} = \frac{1}{\sum_{x} T_x} \sum_{x \in \mathcal{D}_\text{slow}} \sum_{t=1}^{T_x} o_{\ell,h}(x,t),
\end{equation}
and overwrite each ablated head's output with this constant at every evaluation position, for every sample. This removes the head's input-dependent signal while preserving its average contribution to the residual stream; we avoid zero- or uniform-attention substitution, both of which collapse generation entirely.

Tables~\ref{tab:causal_validation_nll} and~\ref{tab:causal_validation_jsd} report the resulting NLL and JSD ratios (ablated vs.\ a size-matched random-head baseline, averaged over three random draws) on a held-out set for both signals. Both effects confirm the respective head sets are causally important for the model's reasoning, despite selecting largely disjoint heads (Table~\ref{tab:independence_stats}).

\begin{table}[h]
\centering
\caption{Causal validation using teacher-forced NLL. $\Delta$NLL denotes the increase from the corresponding unablated baseline under top-5\% signal-selected head ablation or size-matched random-head ablation averaged over three draws. Ratio $=\Delta\mathrm{NLL}_{\mathrm{selected}}/\Delta\mathrm{NLL}_{\mathrm{random}}$. Ratios are computed from unrounded values.}
\label{tab:causal_validation_nll}
\small
\begin{tabular}{llccc}
\toprule
Signal & Domain & $\Delta$NLL$_{\mathrm{selected}}$ & $\Delta$NLL$_{\mathrm{random}}$ & Ratio \\
\midrule
\multirow{3}{*}{Retrieval criticality}
& Music & 0.324 & 0.028 & 11.57$\times$ \\
& Book & 0.562 & 0.114 & 4.93$\times$ \\
& Yelp & 0.341 & 0.046 & 7.35$\times$ \\
\midrule
\multirow{3}{*}{Decision faithfulness}
& Music & 0.751 & 0.078 & 9.65$\times$ \\
& Book & 0.861 & 0.077 & 11.16$\times$ \\
& Yelp & 0.623 & 0.094 & 6.63$\times$ \\
\bottomrule
\end{tabular}
\end{table}

\begin{table}[h]
\centering
\caption{Causal validation (JSD), same setup as Table~\ref{tab:causal_validation_nll}. Ratio $=\text{critical}/\text{random}$.}
\label{tab:causal_validation_jsd}
\small
\begin{tabular}{llccc}
\toprule
Signal & Domain & Critical & Random & Ratio \\
\midrule
\multirow{3}{*}{$R_{\ell,h}$-critical} & Music & 0.119 & 0.036 & 3.29$\times$ \\
 & Book & 0.143 & 0.036 & 3.96$\times$ \\
 & Yelp & 0.108 & 0.034 & 3.19$\times$ \\
\midrule
\multirow{3}{*}{Faith-critical} & Music & 0.154 & 0.027 & 5.64$\times$ \\
 & Book & 0.159 & 0.031 & 5.20$\times$ \\
 & Yelp & 0.121 & 0.029 & 4.19$\times$ \\
\bottomrule
\end{tabular}
\vspace{1em}
\end{table}

\section{Method Design Justifications and Proofs}
\label{app:signal_justifications}

\subsection{Token Weighting and Filtering for Retrieval Criticality}
\label{app:tfidf_weighting}
For each successful retrieval contributing to $R_{\ell,h}$ (\S\ref{sec:head_risk}), we weight the retrieved token $w_t$ by
$\tau_x(w_t)=\mathrm{tf}_x(w_t)\mathrm{idf}(w_t)$, measuring its informativeness within the sample's source text. This gives greater credit to rare, content-bearing tokens that distinguish a particular user or item, while downweighting common tokens that appear across many calibration samples. The following paragraphs define the term-frequency and inverse-document-frequency components.

\paragraph{Stopword and punctuation exclusion.} Decoding steps corresponding to stopwords or punctuation tokens are excluded from Cond.1 and Cond.2 entirely, before the retrieval check is evaluated --- not merely down-weighted --- since such tokens trivially satisfy the retrieval criterion almost anywhere in the evidence region and would otherwise inflate $R_{\ell,h}$ with content-free successes.

\paragraph{Tokenisation and vocabulary.}
Both $\mathrm{tf}$ and $\mathrm{df}$ are computed over the model's subword vocabulary rather than whitespace-delimited words. Each source text is tokenised once, and every resulting token ID is treated as a distinct term. A token is excluded from both statistics if its decoded form either contains no alphanumeric character or, after lowercasing, appears in the NLTK English stopword list~\cite{bird2009natural} --- the same criterion used to filter retrieval events and the evidence-region tokens in $|K^{(s)}(x)|$.

\paragraph{Term frequency.}
For each sample $x$, we define
\[
\mathrm{tf}_x(w)
=
\frac{\mathrm{count}(w,\mathrm{doc}(x))}
{\sum_{w'}\mathrm{count}(w',\mathrm{doc}(x))},
\]
where $\mathrm{doc}(x)$ concatenates the user-history and target-item text fields, excluding the separate average-rating fields attached to each (\texttt{user\_avg\_rating}, \texttt{target\_item\_avg\_rating}), and the denominator counts all retained content tokens.

\paragraph{Inverse document frequency.}
We compute inverse document frequency over the calibration set $\mathcal{D}_{\mathrm{slow}}$, treating each sample's document $\mathrm{doc}(x)$ as one document in the corpus:
\[
\mathrm{idf}(w)
=
\log\frac{1+N}{1+\mathrm{df}(w)}+1,
\]
where $N=|\mathcal{D}_{\mathrm{slow}}|$ and $\mathrm{df}(w)$ is the number of samples' documents containing $w$. Smoothing the numerator and denominator and adding $1$ outside the logarithm keeps the weight finite and positive for both rare and universally occurring tokens.

\subsection{Segment-Level Predictive Importance}
\label{app:segment_jsd}

We anchor decision faithfulness to \textsc{match} because this segment is where user- and item-side evidence is synthesised into a compatibility judgement. If the \textsc{rate} decision is grounded in the reasoning trace, it should rely on this judgement rather than bypassing it and returning to earlier evidence or prompt cues.

We validate this choice by measuring how much each reasoning segment affects the rate-token distribution. For each segment $s \in \{\textsc{user}, \textsc{item}, \textsc{match}\}$, we replace all tokens in that segment with a neutral filler of identical length while keeping the rest of the trace unchanged. We then measure the shift in the output distribution at the \textsc{rate} tokens. A larger shift indicates that the removed segment carries more predictive information for the final decision. We quantify this shift with Jensen--Shannon divergence:
\begin{equation}
    I_s = \frac{1}{|D|}
    \sum_{x\in D}
    \frac{1}{|T_{\textsc{rate}}(x)|}
    \sum_{t\in T_{\textsc{rate}}(x)}
    \mathrm{JSD}\big(p_t(\cdot\mid x) \,\|\, p_t(\cdot\mid x^{(-s)})\big),
    \label{eq:segment_importance}
\end{equation}
where $x^{(-s)}$ denotes the trace with segment $s$ replaced, and $T_{\textsc{rate}}(x)$ denotes the decision-token positions.

\smallskip
Table~\ref{tab:segment_importance} reports $I_s$ with 95\% bootstrap confidence intervals. Across all datasets, \textsc{match} produces the largest distributional shift, 3.3--5.1$\times$ higher than either \textsc{user} or \textsc{item}. Its confidence intervals do not overlap with those of the other segments, indicating that the dominance of \textsc{match} is statistically reliable.

\begin{table}[h]
\centering
\caption{Segment importance $I_s$ measured by mean JSD at the \textsc{rate} tokens, with 95\% bootstrap confidence intervals. The \textsc{match} segment has the strongest effect on the final decision distribution across all datasets.}
\label{tab:segment_importance}
\small
\begin{tabular}{l ccc}
\toprule
\textbf{Dataset} & \textsc{user} & \textsc{item} & \textsc{match} \\
\midrule
\multirow{2}{*}{Book} & 0.0076 & 0.0071 & 0.0287 \\
 & [0.0064, 0.0089] & [0.0062, 0.0080] & [0.0245, 0.0330] \\
\addlinespace
\multirow{2}{*}{Music} & 0.0101 & 0.0077 & 0.0520 \\
 & [0.0086, 0.0117] & [0.0065, 0.0089] & [0.0448, 0.0594] \\
\addlinespace
\multirow{2}{*}{Yelp} & 0.0107 & 0.0118 & 0.0392 \\
 & [0.0097, 0.0117] & [0.0109, 0.0128] & [0.0358, 0.0429] \\
\bottomrule
\end{tabular}
\end{table}

The analysis shows that \textsc{match} is the most direct segment-level carrier of predictive information for the final decision, consistent with its role as the compatibility-synthesis step.

\subsection{Length Sensitivity of Decision Faithfulness}
\label{app:faith_length_bias}

Decision faithfulness uses total attention mass, not per-token density, because the signal is meant to measure how much \textsc{match} evidence enters the decision-time representation, not how much attention \textsc{match} receives per token. We verify that this mass-based definition is robust to region-length sensitivity.

\subsubsection{Length-normalised variants}
\label{par:length-normalised-faith}
Our main definition uses total attention mass because decision faithfulness is intended to measure evidence contribution to the decision representation. A length-normalised alternative instead measures attention per token. Since \textsc{prompt}, \textsc{analyze}, and \textsc{match} differ substantially in length, we evaluate this alternative definition directly.

For each region $r$, let $|\Omega_r(x,t)|$ be the number of positions in that region for trace $x$ and decision token $t$. Density-faith first normalises the region mass at each decision token by the corresponding region length, then averages over tokens and calibration traces:
\begin{equation}
m_{\ell,h}^{r,\mathrm{dens}}
=
\frac{1}{|\mathcal{D}_{\mathrm{slow}}|}
\sum_{x \in \mathcal{D}_{\mathrm{slow}}}
\frac{1}{|T_{\textsc{rate}}(x)|}
\sum_{t \in T_{\textsc{rate}}(x)}
\frac{
\sum_{\tau \in \Omega_r(x,t)}
\mathrm{Att}_{\ell,h}^{(x)}[t,\tau]
}{
|\Omega_r(x,t)|
}.
\end{equation}
This uses each sample's own region length, rather than a fixed dataset-level length. The density-based faithfulness score is
\begin{equation}
\mathrm{faith}^{\mathrm{dens}}_{\ell,h}
=
\frac{
m_{\ell,h}^{\textsc{match},\mathrm{dens}}
}{
m_{\ell,h}^{\textsc{prompt},\mathrm{dens}}
+
m_{\ell,h}^{\textsc{analyze},\mathrm{dens}}
+
m_{\ell,h}^{\textsc{match},\mathrm{dens}}
+
\delta
}.
\end{equation}
The $\sqrt{n}$-faith variant follows the same construction, replacing $|\Omega_r(x,t)|$ with $\sqrt{|\Omega_r(x,t)|}$ before forming the analogous ratio $\mathrm{faith}^{\sqrt{n}}_{\ell,h}$.

\begin{table}[h]
\centering
\caption{Length-sensitivity robustness of decision faithfulness on a matched held-out subset. Density-faith normalises attention mass by region length, while $\sqrt{n}$-faith applies a weaker correction.}
\label{tab:faith_length_robustness}
\small
\begin{tabular}{llcc}
\toprule
Domain & Variant & MAE & RMSE \\
\midrule
\multirow{3}{*}{Yelp}
& Mass-based faithfulness & 0.7513 & 1.0685 \\
& Density-faith & 0.7544 & 1.0751 \\
& $\sqrt{n}$-faith & 0.7575 & 1.0785 \\
\midrule
\multirow{3}{*}{Music}
& Mass-based faithfulness & 0.5292 & 0.8255 \\
& $\sqrt{n}$-faith & 0.5308 & 0.8317 \\
& Density-faith & 0.5311 & 0.8335 \\
\bottomrule
\end{tabular}
\end{table}

Table~\ref{tab:faith_length_robustness} shows that mass-based faithfulness achieves the lowest MAE and RMSE on both datasets. Length-normalised variants change head rankings but do not improve downstream performance.

\subsubsection{Uniform-attention baseline and allocation invariance}
\label{par:baseline-invariance}
\paragraph{Uniform-attention baseline.}
A head with no genuine regional preference would still assign nonzero attention to \textsc{match}, since \textsc{match} occupies part of the visible context. Let $m_0^r$ denote the attention mass that region $r$ would receive under uniform attention, which depends only on region length. The corresponding baseline for decision faithfulness is
\begin{equation}
\eta
=
\frac{
m_0^{\textsc{match}}
}{
m_0^{\textsc{prompt}}
+
m_0^{\textsc{analyze}}
+
m_0^{\textsc{match}}
}.
\end{equation}
This value is identical across heads because it depends only on region sizes, not on head behaviour. In our calibration traces, $\eta \approx 0.036$ on Music and $\eta \approx 0.048$ on Yelp.

\paragraph{Allocation invariance of baseline subtraction.}
The allocation is invariant to any constant shift applied to $\mathrm{faith}_b$, which we use here to show that subtracting the uniform-attention baseline $\eta$ has no effect on the merged model. For a shift $\widetilde{\mathrm{faith}}_b = \mathrm{faith}_b - \eta$, substituting into the perturbation weight gives
\begin{equation}
\begin{aligned}
\widetilde s_b
&=
\exp\!\left(\gamma(\kappa_b + \mathrm{faith}_b - \eta)\right) d_b \\
&=
\exp\!\left(\gamma(\kappa_b + \mathrm{faith}_b)\right) e^{-\gamma\eta} d_b
=
e^{-\gamma\eta} s_b.
\end{aligned}
\end{equation}
Every block's perturbation weight is rescaled by the same constant $e^{-\gamma\eta}$, independent of $b$. Since the injection budget is a fixed fraction of the total perturbation weight,
\begin{equation}
\varepsilon
=
\rho\,\bar\alpha^2
\sum_{b\in\mathcal{B}} s_b,
\end{equation}
the budget rescales by the same factor, so the water-filling solution $\alpha_b^\star$ is unchanged before and after the shift. This argument holds for any constant $\eta$ and any definition of $\mathrm{faith}_b$, so recentering the faithfulness signal --- for the uniform-attention baseline or otherwise --- does not change the merged model, supporting total decision-time attention mass as the operational measure of decision faithfulness.

\subsection{Derivation of the Fisher-Weighted Update-Sensitivity Surrogate}
\label{app:fisher_derivation}

\subsubsection{Local Quadratic Expansion}
\label{app:fisher_derivation_quadratic}
In \S\ref{sec:head_update_sensitivity}, the teacher-forced loss is computed using the slow-thinking model parameters $\theta_S$. To examine how this loss changes as the parameters are perturbed during merging, we denote by $\ell_{\theta}(x)$ the teacher-forced loss computed using an arbitrary parameter vector $\theta$:
\begin{equation}
\ell_{\theta}(x) = -\frac{1}{|Y(x)|} \sum_{t=1}^{|Y(x)|} \log p_{\theta}\!\left(y_t \mid y_{<t},\mathrm{prompt}(x)\right).
\label{eq:app_general_loss}
\end{equation}
Setting $\theta=\theta_S$ recovers the loss defined in \S\ref{sec:head_update_sensitivity}. We average this loss over the calibration set:
\begin{equation}
\bar{\ell}(\theta) = \frac{1}{|\mathcal{D}_{\mathrm{slow}}|} \sum_{x\in\mathcal{D}_{\mathrm{slow}}} \ell_{\theta}(x).
\label{eq:app_average_loss}
\end{equation}
Let $b$ index a merge unit: either the paired $Q/O$ parameters
of a query head or the paired $K/V$ parameters of a GQA group,
and let $\mathcal{B}$ denote the set of all such units. For each
$b\in\mathcal{B}$, let $\Delta_{F,b}$ denote the fast-thinking
update restricted to all parameters belonging to unit $b$.

Given merge coefficients
$\{\alpha_b\}_{b\in\mathcal{B}}$, the total parameter change
applied to the slow-thinking model is
\[
\Delta_\theta(\boldsymbol{\alpha})
=
\sum_{b\in\mathcal{B}}
\alpha_b\Delta_{F,b},
\]
where $\boldsymbol{\alpha}$ collects the coefficients assigned to
all merge units. The resulting merged model is
\[
\theta(\boldsymbol{\alpha})
=
\theta_S+\Delta_\theta(\boldsymbol{\alpha}).
\]
Full injection of the fast-thinking update corresponds to
$\alpha_b=1$ for every merge unit $b$.

To assess how the merged model $\theta(\boldsymbol{\alpha})$ perturbs the calibration loss, we expand $\bar{\ell}$ around $\theta_S$. A second-order Taylor expansion gives
\begin{equation}
\begin{aligned}
\bar{\ell}\!\left(\theta(\boldsymbol{\alpha})\right)
-
\bar{\ell}(\theta_S)
&=
\nabla_{\theta}\bar{\ell}(\theta_S)^{\top}
\Delta_\theta(\boldsymbol{\alpha}) \\
&\quad+
\frac{1}{2}
\Delta_\theta(\boldsymbol{\alpha})^{\top}
H_S
\Delta_\theta(\boldsymbol{\alpha})
+
\mathcal{O}\!\left(
\|\Delta_\theta(\boldsymbol{\alpha})\|_2^3
\right),
\end{aligned}
\label{eq:app_taylor}
\end{equation}
where $H_S=\nabla_{\theta}^{2}\bar{\ell}(\theta_S)$ is the Hessian of the calibration loss at the slow-thinking model. Although $\theta_S$ is already trained, the gradient $\nabla_\theta\bar{\ell}(\theta_S)$ need not vanish because $\mathcal{D}_{\mathrm{slow}}$ is a filtered calibration set rather than the original training set. We omit the first-order term because it is signed and can be negative or cancel across merge units depending on the alignment between $\Delta_\theta(\boldsymbol{\alpha})$ and the calibration gradient, making it unsuitable as a non-negative perturbation cost. By contrast, under a positive-semidefinite (PSD) curvature approximation, the quadratic term is non-negative and measures the local curvature along the merge-update direction. We therefore use it as a non-negative local sensitivity surrogate:
\begin{equation}
C_H(\boldsymbol\alpha) = \frac{1}{2} \Delta_\theta(\boldsymbol\alpha)^{\top} H_S \Delta_\theta(\boldsymbol\alpha).
\label{eq:app_hessian_cost}
\end{equation}
This quantity captures the local curvature of the calibration loss along the direction of the merge-induced parameter change. A larger value signals that the fast-thinking update perturbs the slow-thinking model along a direction of higher loss curvature, and therefore carries greater risk to model behaviour.

\subsubsection{From Hessian to a Diagonal Empirical-Fisher Surrogate}

\paragraph{Hessian decomposition.}
For a negative log-likelihood objective, the Hessian contains two sources of curvature, following the generalised Gauss--Newton decomposition of~\cite{martens2020newinsightsperspectivesnatural}. The first captures how the loss curves with respect to the model logits and how parameter changes propagate to those logits; this forms the generalised Gauss--Newton component. The second arises from the nonlinear dependence of the logits on the model parameters and forms a residual term.

For each example $x$, let $z_{\theta}(x)$ denote the vector obtained by stacking the logits used at all teacher-forced response positions, and let $J_{\theta}(x)=\partial z_{\theta}(x)/\partial \theta$ denote its Jacobian with respect to the model parameters. Applying the chain rule twice and averaging over the calibration set gives
\begin{equation}
\begin{aligned}
H_S
&=
\nabla_{\theta}^{2}\bar{\ell}(\theta_S)
\\
&=
\frac{1}{|\mathcal{D}_{\mathrm{slow}}|}
\sum_{x\in\mathcal{D}_{\mathrm{slow}}}
\Bigg[
\underbrace{
J_{\theta_S}(x)^{\top}
\left.
\nabla_z^2\ell(z;x)
\right|_{z=z_{\theta_S}(x)}
J_{\theta_S}(x)
}_{G_S(x)}
\\
&\qquad\qquad\qquad+
\underbrace{
\sum_k
\left.
\frac{\partial \ell(z;x)}{\partial z_k}
\right|_{z=z_{\theta_S}(x)}
\left.
\nabla_{\theta}^{2}z_{\theta,k}(x)
\right|_{\theta=\theta_S}
}_{R_S(x)}
\Bigg]
\\
&=
G_S+R_S,
\end{aligned}
\label{eq:app_hessian_decomposition}
\end{equation}
where the index $k$ runs over all scalar entries of the stacked logit vector, and $G_S=\frac{1}{|\mathcal{D}_{\mathrm{slow}}|}\sum_x G_S(x)$, $R_S=\frac{1}{|\mathcal{D}_{\mathrm{slow}}|}\sum_x R_S(x)$.

\paragraph{Model Fisher.}
We next relate $G_S$ to the Fisher information matrix, which measures how sensitive the model's predicted next-token distribution is to changes in its parameters. At teacher-forced decoding position $t$ of example $x$, let $c_t(x)=(\mathrm{prompt}(x),y_{<t})$ denote the fixed context preceding that position. Given this context, the model produces a logit $z_{\theta,k}(x,t)$ for each vocabulary token $k$, and the softmax function converts these logits into the next-token distribution
\[
p_{\theta}\!\left(y\mid c_t(x)\right)
=
\frac{
\exp\!\left(z_{\theta,y}(x,t)\right)
}{
\sum_{k\in\mathcal{V}}
\exp\!\left(z_{\theta,k}(x,t)\right)
},
\]
where $\mathcal{V}$ denotes the vocabulary. Let $J_{\theta}(x,t)=\partial z_{\theta}(x,t)/\partial\theta$ denote the Jacobian of the logit vector at position $t$, and define the score function $s_{\theta}(y,x,t)=\nabla_{\theta}\log p_{\theta}(y\mid c_t(x))$. The token-level conditional model Fisher at this position is
\begin{align*}
\mathcal{F}_S(x,t)
&=
\mathbb{E}_{y\sim p_{\theta_S}(\cdot\mid c_t(x))}
\left[
s_{\theta_S}(y,x,t)
s_{\theta_S}(y,x,t)^{\top}
\right]
\\
&=
\sum_{y\in\mathcal{V}}
p_{\theta_S}\!\left(y\mid c_t(x)\right)
s_{\theta_S}(y,x,t)
s_{\theta_S}(y,x,t)^{\top},
\end{align*}
where the second form follows because the output vocabulary is discrete. Thus, the model Fisher considers all possible next tokens, weighted by the probabilities assigned to them by the slow-thinking model, rather than using only the observed target token $y_t$. Averaging over the same calibration examples and response positions as the teacher-forced loss gives
\begin{equation}
\mathcal{F}_S
=
\frac{1}{|\mathcal{D}_{\mathrm{slow}}|}
\sum_{x\in\mathcal{D}_{\mathrm{slow}}}
\frac{1}{|Y(x)|}
\sum_{t=1}^{|Y(x)|}
\mathcal{F}_S(x,t).
\label{eq:app_model_fisher}
\end{equation}

\paragraph{Gauss--Newton equals Fisher.}
We next establish that, for categorical negative log-likelihood with softmax logits, the model Fisher matrix $\mathcal{F}_S$ coincides exactly with the generalised Gauss--Newton matrix $G_S$~\cite{martens2020newinsightsperspectivesnatural}. Consider a fixed example $x$ and decoding position $t$, and treat the per-position loss as a function of a free logit vector $z$. Let $p=p_{\theta}(\cdot\mid c_t(x))$ denote the corresponding softmax distribution, and let $\ell_t(z)=-\log p_y$ be the loss for observed token $y$. The Hessian of this loss with respect to the logits is
\[
\nabla_z^2\ell_t(z)=\operatorname{diag}(p)-pp^{\top},
\]
which is independent of the observed token. Meanwhile, the log-probability score is
\[
\nabla_z\log p(y)=e_y-p,
\]
where $e_y$ denotes the one-hot vector for token $y$. Taking its expected outer product under $y\sim p$ yields
\begin{align*}
\mathbb{E}_{y\sim p}
\left[
(e_y-p)(e_y-p)^{\top}
\right]
&=
\mathbb{E}_{y\sim p}
\left[
e_ye_y^{\top}
\right]
-
pp^{\top}
\\
&=
\operatorname{diag}(p)-pp^{\top}.
\end{align*}
The logit-space curvature of the loss therefore equals the covariance of the logit-space score. Mapping both quantities into parameter space through $J_{\theta_S}(x,t)$ gives $G_S(x,t)=J_{\theta_S}(x,t)^{\top}[\operatorname{diag}(p)-pp^{\top}]J_{\theta_S}(x,t)=\mathcal{F}_S(x,t)$. Moreover, because the teacher-forced loss averages the negative log-likelihood over response positions, $G_S(x)=\frac{1}{|Y(x)|}\sum_{t=1}^{|Y(x)|}G_S(x,t)$, and averaging over the calibration set therefore yields $G_S=\mathcal{F}_S$.

\paragraph{Discarding the residual.}
Following the standard generalised Gauss--Newton approximation, we discard the residual curvature term $R_S$: near a well-trained solution such as $\theta_S$, the per-token prediction error entering $R_S$ is small, so its contribution to the Hessian is expected to be small relative to $G_S$. The Hessian is then approximated by the positive-semidefinite model Fisher:
\begin{equation}
H_S
=
G_S+R_S
\approx
G_S
=
\mathcal{F}_S.
\label{eq:app_hessian_fisher}
\end{equation}

\paragraph{Empirical-Fisher approximation.}
Computing the model Fisher $\mathcal{F}_S$ exactly would require evaluating the score for every possible vocabulary token at every response position. This is infeasible for a large language model. We therefore replace it with the empirical Fisher, which uses the observed teacher-forced responses in the calibration set.

For each example $x$, define the gradient of its teacher-forced loss at the slow-thinking parameters as $g(x)=\left.\nabla_{\theta}\ell_{\theta}(x)\right|_{\theta=\theta_S}$. The empirical Fisher is the average outer product of these gradients:
\begin{equation}
\widehat{\mathcal{F}}_{\mathrm{emp}}
=
\frac{1}{|\mathcal{D}_{\mathrm{slow}}|}
\sum_{x\in\mathcal{D}_{\mathrm{slow}}}
g(x)g(x)^{\top}.
\label{eq:app_empirical_fisher}
\end{equation}
Unlike the model Fisher in Eq.~\eqref{eq:app_model_fisher}, this estimator uses the observed response $Y(x)$ rather than averaging over tokens drawn from the model's predictive distribution. It is therefore not exactly equal to either $\mathcal{F}_S$ or $H_S$, but provides a tractable positive-semidefinite curvature surrogate~\cite{kunstner2020limitationsempiricalfisherapproximation}.

Storing the full empirical-Fisher matrix remains infeasible because its size grows quadratically with the number of model parameters. We therefore retain only its diagonal entries, $\widehat{\mathcal{F}}_{\mathrm{emp}}\approx\operatorname{diag}(F_1,\ldots,F_{|\theta|})$, where
\[
F_j
=
\frac{1}{|\mathcal{D}_{\mathrm{slow}}|}
\sum_{x\in\mathcal{D}_{\mathrm{slow}}}
\left(
\left.
\frac{\partial \ell_{\theta}(x)}
{\partial \theta_j}
\right|_{\theta=\theta_S}
\right)^2.
\]
Each $F_j$ measures how strongly the calibration loss responds locally to changes in parameter $\theta_j$. A larger value indicates that a small change to that parameter is more likely to disturb the slow-thinking behaviour captured by the calibration set.

\subsection{Derivation of the Head-Level Update Sensitivity}
\label{app:head_update_sensitivity_derivation}

We use the diagonal empirical Fisher to approximate the curvature cost of the merged parameter change $\Delta_\theta(\boldsymbol{\alpha})=\sum_{b\in\mathcal{B}}\alpha_b\Delta_{F,b}$, where $b$ indexes either a paired $Q/O$ head or a paired $K/V$ group. The resulting sensitivity surrogate is
\[
C_F(\boldsymbol{\alpha})
=
\frac{1}{2}
\Delta_\theta(\boldsymbol{\alpha})^{\top}
\operatorname{diag}(F_1,\ldots,F_{|\theta|})
\Delta_\theta(\boldsymbol{\alpha})
=
\frac{1}{2}
\sum_j
F_j
\left(\Delta_{\theta,j}(\boldsymbol{\alpha})\right)^2.
\]

Each merge unit $b$ comprises two individual projection slices, denoted by $c\in\mathcal{C}(b)$: the $Q$ and $O$ slices of a query head, or the $K$ and $V$ slices of a GQA group. Both slices receive the same coefficient $\alpha_b$. For each slice, its update sensitivity is $d_c=\sum_{j\in c}F_j(\Delta_{F,j})^2$, as defined in Eq.~\eqref{eq:fisher_sensitivity}. Because the projection slices are disjoint, substituting the merge-induced update gives
\begin{equation}
\begin{aligned}
C_F(\boldsymbol{\alpha})
&=
\frac{1}{2}
\sum_{b\in\mathcal{B}}
\alpha_b^2
\sum_{c\in\mathcal{C}(b)}
d_c \\
&=
\sum_{b\in\mathcal{B}}
\alpha_b^2 d_b,
\end{aligned}
\label{eq:app_block_fisher_cost}
\end{equation}
where $d_b=\frac{1}{2}\sum_{c\in\mathcal{C}(b)}d_c$ is the averaged sensitivity of the two projection slices in merge unit $b$. This corresponds to $d_{\ell,h}^{QO}=(d_{\ell,h}^{Q}+d_{\ell,h}^{O})/2$ for a query head and $d_{\ell,g}^{KV}=(d_{\ell,g}^{K}+d_{\ell,g}^{V})/2$ for a GQA group, matching the definition in \S\ref{sec:head_update_sensitivity}.

Thus, $d_b$ combines the sensitivity of the slow-thinking model with the magnitude of the fast-thinking update for merge unit $b$. A larger $d_b$ indicates that the proposed update is more likely to disturb the behaviour captured by the calibration loss. Its contribution to the curvature surrogate scales quadratically with $\alpha_b$, motivating the perturbation constraint in \S\ref{sec:allocation}.

\subsection{Form of the Perturbation-Cost Multiplier}
\label{app:s_b_form}

We ablate the link function used to map retrieval criticality and decision faithfulness into a multiplier on Fisher risk. The main method uses
\[
s_b
=
m_b d_b,
\qquad
m_b
=
\exp\!\left(\gamma(\kappa_b+\mathrm{faith}_b)\right).
\]
We compare this exponential multiplier with two alternatives using the same signals:
\[
\begin{aligned}
m_b^{\mathrm{lin}}
&=
\max\!\left(0.05,\ 1+\gamma(\kappa_b+\mathrm{faith}_b)\right),\\
m_b^{\mathrm{sig}}
&=
1+\gamma\!\left(\sigma(z(\kappa_b))+\sigma(z(\mathrm{faith}_b))\right).
\end{aligned}
\]
Here $z(x)=(x-\bar{x})/s_x$ standardises each signal across heads, and $\sigma(x)=1/(1+e^{-x})$ maps it to $(0,1)$. Since $\kappa_b$ and $\mathrm{faith}_b$ are already bounded in $[0,1]$, this variant stretches relative differences by z-scoring, then compresses them through the sigmoid. The linear floor is only a numerical safeguard for near-zero denominators in the water-filling solution.

The three forms differ mainly in dynamic range: the exponential form is multiplicative and unbounded, so blocks with jointly high retrieval criticality and decision faithfulness receive substantially larger perturbation weights than either signal alone would produce.

\begin{table}[h]
\centering
\caption{Dynamic range of the perturbation multiplier under each link function, measured across all heads on Music.}
\label{tab:link_range}
\small
\begin{tabular}{lccc}
\toprule
Link function & Min & Max & Ratio \\
\midrule
Exponential & 2.3 & 108 & $\sim$47$\times$ \\
Linear & 1.8 & 5.7 & $\sim$3.2$\times$ \\
Sigmoid & 2.7 & 6.3 & $\sim$2.3$\times$ \\
\bottomrule
\end{tabular}
\end{table}

As shown in Table~\ref{tab:link_range}, the exponential form produces a substantially wider spread of perturbation weights than the linear and sigmoid alternatives. This allows the allocation to separate high-cost blocks more sharply, whereas the alternatives compress most blocks into a narrower range.

\begin{table}[h]
\centering
\caption{Perturbation-multiplier ablation on Yelp ($n=400$).}
\label{tab:s_b_form}
\small
\begin{tabular}{lcc}
\toprule
Link function & MAE & RMSE \\
\midrule
Exponential & \textbf{0.7348} & \textbf{1.0656} \\
Sigmoid & 0.7617 & 1.0970 \\
Linear & 0.7686 & 1.1030 \\
\bottomrule
\end{tabular}
\end{table}

Table~\ref{tab:s_b_form} compares the three multiplier forms on
the same matched Yelp subset. The exponential form achieves the lowest MAE and RMSE, supporting its use in the main method. Because this ablation uses a smaller matched subset, the values are reported separately from the full test-set production results.

\subsection{Normalisation Choices in the Perturbation Cost}
\label{app:normalization_choices}

The perturbation cost
\[
s_b=\exp(\gamma(\kappa_b+\mathrm{faith}_b))\,d_b
\]
combines three quantities with different meanings and scales. We therefore normalise them differently rather than applying a uniform preprocessing rule.

\paragraph{Retrieval criticality.}
The raw retrieval criticality $R_{\ell,h}$ has no natural scale and is highly skewed. Since retrieval criticality enters the exponent, using raw $R_{\ell,h}$ would allow extreme-valued heads to dominate the allocation. We therefore map $R_{\ell,h}$ through the log--z-score--sigmoid transform defined in \S\ref{sec:method}, producing a bounded and comparable value $\kappa_b$ before exponentiation.

\paragraph{Decision faithfulness.}
We use $\mathrm{faith}_b$ in its raw form. Unlike $R_{\ell,h}$, decision faithfulness is already bounded and directly interpretable as the share of tracked decision-time attention directed to \textsc{match}. We therefore apply no additional transformation.

\paragraph{Fisher update sensitivity.}
We also retain $d_b$ on its original Fisher-weighted scale. Unlike $\kappa_b$ and $\mathrm{faith}_b$, $d_b$ is not a relative score but an estimate of update-induced perturbation: loss sensitivity weighted by the squared update magnitude. Nonlinear or rank transforms would discard this magnitude and weaken the link between the perturbation constraint and the Fisher-weighted cost. Uniform rescaling leaves the solution unchanged, since it rescales $s_b$ and $\varepsilon$ equally; but transforming relative magnitudes changes the allocation.

\subsection{KKT Derivation of the Water-Filling Solution}
\label{app:kkt_derivation}

The allocation problem in \S\ref{sec:allocation} is
\[
\max_{\alpha}\; \sum_{b\in\mathcal{B}} w_b \alpha_b
\qquad \text{s.t.} \qquad
\sum_{b\in\mathcal{B}} s_b \alpha_b^2 \le \varepsilon,
\qquad 0 \le \alpha_b \le \bar\alpha .
\]
Since the objective is linear and the feasible set is convex when $s_b\ge0$, the KKT conditions are sufficient for global optimality.

The coefficients $\alpha_b$ share a single perturbation budget $\varepsilon$, so they cannot be chosen independently. We introduce a multiplier $\mu\ge0$ for the shared constraint; for fixed $\mu$, the problem separates across blocks, and $\mu$ is later adjusted until the resulting coefficients satisfy the budget condition. The Lagrangian objective is
\[
\sum_{b\in\mathcal{B}}
\left(
w_b\alpha_b
-
\mu s_b\alpha_b^2
\right)
+
\mu\varepsilon .
\]
Thus each coordinate solves
\[
\max_{0\le \alpha_b\le \bar\alpha}
\; w_b\alpha_b-\mu s_b\alpha_b^2 .
\]
For $s_b>0$ and $\mu>0$, the unconstrained stationary point satisfies
\[
\frac{\partial}{\partial \alpha_b}
\left(
w_b\alpha_b-\mu s_b\alpha_b^2
\right)
=
w_b-2\mu s_b\alpha_b
=
0,
\qquad
\alpha_b(\mu)
=
\frac{w_b}{2\mu s_b}.
\]
Applying the box constraint gives the water-filling solution
\[
\alpha_b^\star
=
\operatorname{clip}\!\left(
\frac{w_b}{2\mu s_b},
0,
\bar\alpha
\right).
\]
The multiplier $\mu$ is not a fixed hyperparameter; it is solved separately for each merge. In the binding case, $\mu>0$ is found by bisection
\[
\sum_{b\in\mathcal{B}} s_b(\alpha_b^\star)^2
=
\varepsilon .
\]
If full injection already satisfies the perturbation constraint, then the constraint is slack. Complementary slackness sets $\mu=0$, and the optimal solution is $\alpha_b^\star=\bar\alpha$ for all blocks.

When $s_b=0$, block $b$ carries no perturbation penalty. If $w_b>0$, the objective is maximised by setting $\alpha_b^\star=\bar\alpha$; if $w_b=0$, all values in $[0,\bar\alpha]$ are equivalent, and we set $\alpha_b^\star=\bar\alpha$ by convention.

\subsection{Generalisability of FFN Exclusion Across Model Scale and Family}
\label{app:ffn_generalization}
\begin{figure}[h]
    \centering
    \includegraphics[width=\columnwidth]{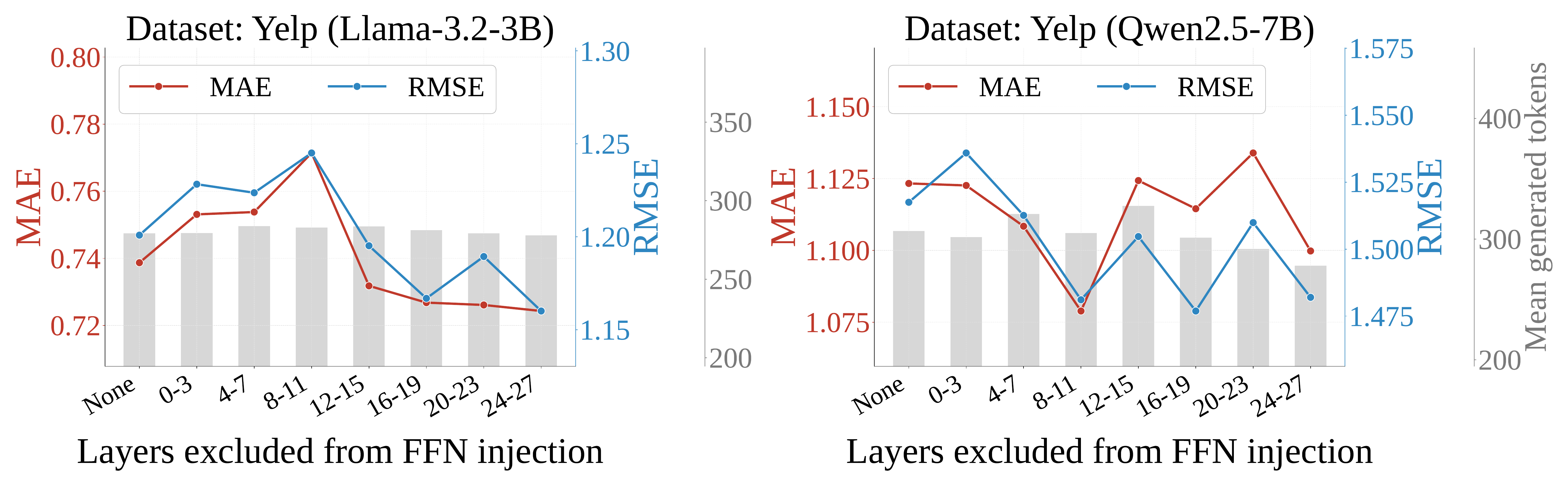}
    \caption{FFN exclusion sweep on Yelp for Llama-3.2-3B-Instruct (left) and Qwen2.5-7B-Instruct (right), both using four-layer exclusion windows. Bars show mean generated tokens; lines show MAE and RMSE.}
    \label{fig:ffn_generalization}
\end{figure}

To assess late-layer FFN exclusion beyond the 3B setup in \S\ref{sec:rq3-ffn}, we repeat the Yelp sweep using Llama-3.2-3B-Instruct and Qwen2.5-7B-Instruct (Figure~\ref{fig:ffn_generalization}). Both contain 28 layers, compared with 36 in Qwen2.5-3B. We therefore exclude four-layer windows to preserve a comparable fraction of model depth, rather than reusing six-layer windows for the deeper backbone. The window length is therefore scaled to the backbone depth, while
its location is selected by the validation sweep. Once selected,
the resulting window is fixed for all evaluations using that
backbone.

On Llama-3.2-3B, excluding progressively later layers improves both MAE and RMSE, with layers 24--27 yielding the best accuracy--efficiency trade-off. This is consistent with the main-text finding that late FFNs are particularly sensitive to the fast-thinking update. On Qwen2.5-7B, however, the pattern is less monotonic: excluding either layers 8--11 or 24--27 gives competitive accuracy, whereas excluding layers 20--23 produces the worst MAE and RMSE among the tested ranges. These results support excluding late FFN layers as a robust default across model scales and families: doing so consistently improves the accuracy--efficiency trade-off relative to unrestricted merging, even though the single most sensitive range shifts modestly across backbones. Accordingly, late-layer exclusion serves as the default strategy,
but the exact exclusion window should be validated once for each
new backbone rather than transferred directly across architectures.

\end{document}